\documentclass[aps,prb,twocolumn,superscriptaddress]{revtex4-2}

\usepackage[hidelinks]{hyperref}
\hypersetup{
	colorlinks,
	linkcolor={red},
	citecolor={blue},
	urlcolor={blue}
}
\usepackage{physics}
\usepackage{balance}
\usepackage{amssymb}
\usepackage{suffix}
\usepackage{mathtools}
\usepackage[utf8]{inputenc}
\usepackage{booktabs}
\usepackage{cases}
\usepackage[multiple]{footmisc}
\usepackage{dcolumn}
\usepackage{color,soul}
\usepackage{rotating}
\usepackage{perpage}
\usepackage{siunitx}
\usepackage{xcolor}
\usepackage{soul}
\usepackage{amsmath}
\usepackage{tikz}
\usepackage[T1]{fontenc}
\usepackage{etoolbox}
\usepackage{graphics}
\usepackage{siunitx}
\usepackage{float}	
\usepackage{collref}
\usepackage{multirow}
\usepackage{mathtools}
\usepackage{bm}
\usepackage{url}

\usepackage{tikz}
\usepackage{tikz-3dplot}
\usepackage{accents}

\makeatletter
\newcommand{\doublewidetilde}[1]{{%
  \mathpalette\double@widetilde{#1}%
}}
\newcommand{\double@widetilde}[2]{%
  \sbox\z@{$\m@th#1\widetilde{#2}$}%
  \ht\z@=.9\ht\z@
  \widetilde{\box\z@}%
}
\makeatother

\usepackage{etoolbox,lipsum}

\begin{document} 

 \title{Robust qubit interactions mediated by photonic topological edge states}

 \author{Boris Gurevich}
 \email{Contact author: boris.gurevich@utdallas.edu}
 \address{Department of Physics, The University of Texas at Dallas, Richardson, Texas 75080, USA}
 \author{Weihua Xie}
 \address{Department of Physics, The University of Texas at Dallas, Richardson, Texas 75080, USA}
 \author{Mohsen Yarmohammadi}
 \address{Department of Physics, Georgetown University, Washington DC 20057, USA} 
 \author{Michael H. Kolodrubetz}
 \address{Department of Physics, The University of Texas at Dallas, Richardson, Texas 75080, USA}

 \date{\today}
\begin{abstract}

We investigate the coupling of two spatially separated qubits via topologically protected edge states in a two-dimensional Hofstadter lattice. In this hybrid platform, the qubits are coupled to distinct edge sites of the lattice, enabling long-range interactions mediated by topological edge modes. We solve the full system Hamiltonian and analyze the resulting eigenstate structure to uncover the conditions under which coherent qubit interactions emerge. Our analysis reveals that the effective coupling is highly sensitive to the qubit placement, energy detuning, and the topological character of the edge spectrum. We obtain an analytical solution that goes beyond the perturbative regime, capturing the full interplay between the qubits and edge modes. These results provide a foundation for exploring information transport and many-body effects in engineered quantum systems where interactions are mediated by topological edge modes.

\end{abstract}
\maketitle
	{\allowdisplaybreaks
\section{Introduction}

The study of engineered topological states of matter has become a central focus in condensed matter physics, driving progress in both theory and experiment \cite{Kiczynski2022, Zhang02102018, Rachel_2018}. A particularly compelling direction is the exploration of synthetic materials constructed using light, which can exhibit topological properties with promising implications for future technologies \cite{RevModPhys.91.015006, PhysRevLett.127.093901, Bao2022, Tan2014, Rechtsman_2013}. This interdisciplinary field, at the crossroads of quantum many-body physics and photonics, continues to open new avenues for research and applications \cite{RevModPhys.85.299, RevModPhys.91.015006, Krishnamoorthy2023}.

Among the various theoretical models, the Hofstadter model stands out as a simple framework for capturing the essential physics of Chern insulators on a lattice \cite{PhysRevB.28.6713, PhysRevB.98.024205, PhysRevB.90.115132, PhysRevB.90.205111, PhysRevLett.125.236804, PhysRevB.14.2239}. It has gained renewed interest in recent years, particularly due to its experimental realization in superconducting resonator arrays and transmon qubit platforms, which offer access to chiral edge modes and site-resolved spectroscopic measurements \cite{Owens2022, PhysRevA.97.013818}. These experiments, which realize a quarter-flux Hofstadter lattice, have enabled manipulation of individual photonic modes and serve as a powerful platform for probing topological effects in synthetic systems.

One area of growing interest involves the interaction between qubits coupled to such structured lattices \cite{PhysRevB.106.235409, Gu_2019}. In this context, indirect qubit-qubit coupling via topologically protected edge states offers an alternative to conventional coupling schemes, potentially enhancing robustness against noise and parameter fluctuations \cite{PhysRevA.98.012331, Pakkiam2023, doi:10.1126/science.abq5769, PhysRevA.94.013620}. This mechanism bears resemblance to the Ruderman–Kittel–Kasuya–Yosida (RKKY) interaction ~\cite{10.1143/PTP.16.45,* PhysRev.106.893, *PhysRev.96.99},  albeit mediated here by one-way chiral edge modes rather than spin exchange with itinerant bulk modes.

Despite these promising features and the significant progress reported in Refs.~\cite{PhysRevB.106.235409, PhysRevA.108.043708, PhysRevA.105.023329, PhysRevLett.122.010406, PhysRevA.94.043611} on related models, a complete theoretical framework—particularly one that provides explicit solutions for such mediated interactions in the strong coupling regime—remains underdeveloped.

In this work, we present a detailed theoretical analysis of the indirect interaction between two qubits coupled to a topological Hofstadter lattice. We derive analytical results that span a broad range of coupling strengths and include the effects of imperfect calibration. Our approach provides a picture of qubit interactions mediated by topological edge states and offers insights that can guide future experimental designs. By capturing key features such as non-perturbative interactions and robustness to disorder, this framework contributes to the foundational understanding necessary for advancing topologically robust quantum platforms.

The remainder of this paper is organized as follows. In Sec. \ref{s2}, we review the Hofstadter model in the context of synthetic topological systems and introduce the theoretical framework describing qubit coupling via edge states. Sections \ref{sec_perturb} and \ref{sec_finite} present the derivation of the effective qubit interactions and their dependence on system parameters, in the perturbative and non-perturbative regimes, respectively. In Sec. \ref{sec_fidelity}, we provide a detailed analysis and closed-form solution for the oscillation fidelity. \textcolor{black}{Section \ref{sec_experiment} proposes a concrete experimental realization.} Finally, Sec. \ref{sec_concl} concludes with a summary and outlook, discussing the physical implications of our results and highlighting their relevance for robust quantum coupling schemes.

\section{Model}\label{s2}\begin{figure}[t]
	\centering
	\includegraphics[width=0.95\linewidth]{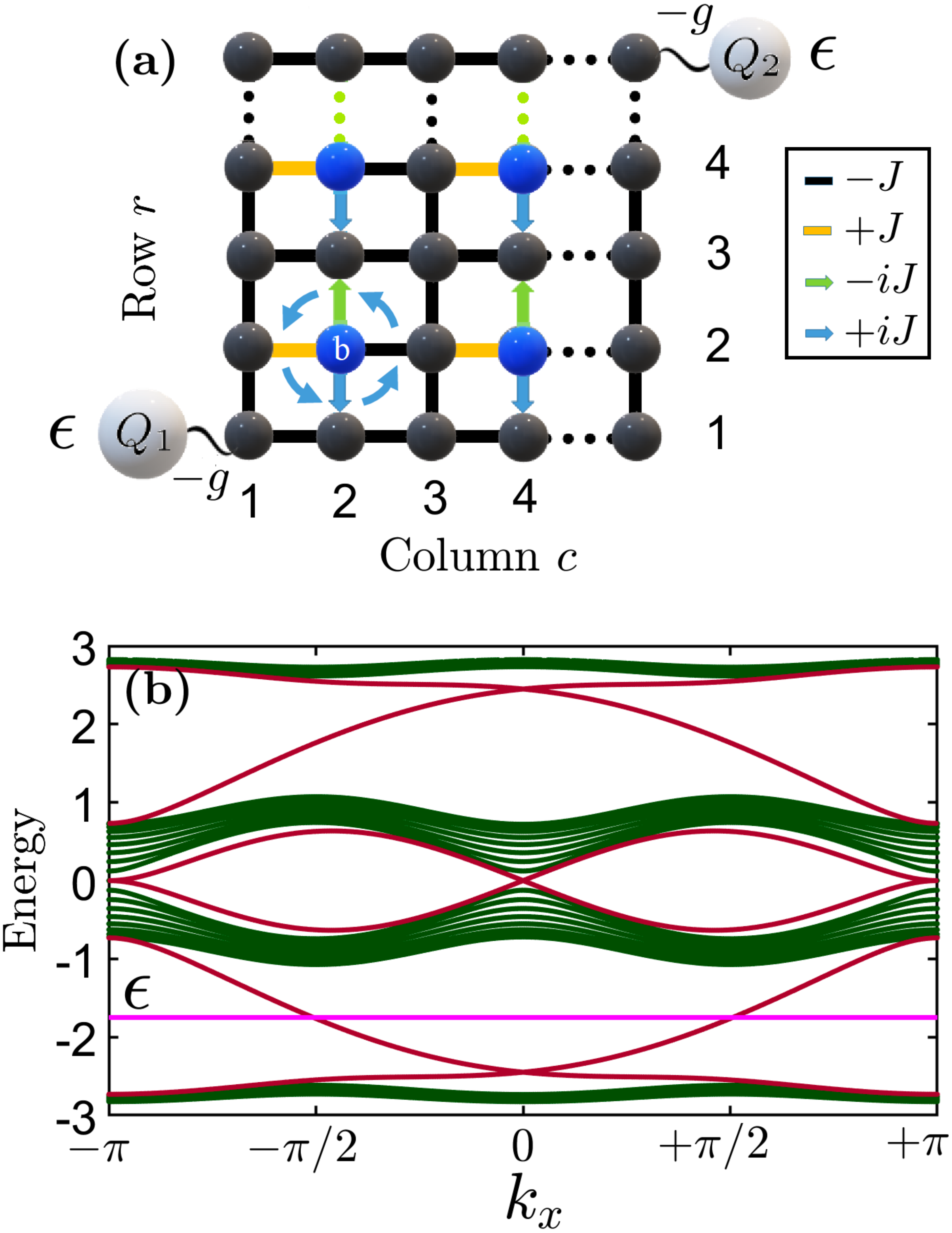}
	\caption{\label{fig:qbt_hofst_lat}(a) Two qubits, \(Q_{1}\) and \(Q_{2}\), are coupled to a Hofstadter lattice (\(L \times L\)) of microwave resonators connected via nearest-neighbor hopping $J$.  Blue sites support \(p_x + ip_y\) modes, while black sites host \(s\)-like modes. Site indices follow \(j(r,c) = (r-1)L + c\), and the full system includes \(N = L^2 + 2\) sites with \(Q_{1} = 0\), \(Q_{2} = L^2 + 1\). A rotating frame is chosen such that the qubits have nonzero on-site potentials \(\epsilon\), while the lattice is centered around zero energy. (b) Band structure for the Hofstadter lattice in a strip geometry (\(L_y = 35\), \(L_x \to \infty\)). Bulk bands \textcolor{black}{(green)} lie in \( |E| \gtrsim 2.61J \) and \( |E| \lesssim 1.08J \); edge states (red) span the intermediate region. The qubit energy \(\epsilon \approx -1.75J\) (\textcolor{black}{magenta}) lies within the edge spectrum.
	}
\end{figure}
Inspired by the seminal experimental results in Refs.~\cite {PhysRevA.97.013818, Owens2022}, we investigate the dynamics of two qubits, $Q_1$ and $Q_2$, each weakly coupled to spatially separated edge sites of an $L\times L$ Hofstadter lattice, as depicted in Fig.~\ref{fig:qbt_hofst_lat}(a).
The lattice sites consist of microwave resonators with frequency $\omega_0$, which interact with adjacent sites via photon tunneling, characterized by an effective hopping amplitude $J$. In a typical configuration~\cite{Owens2022}, $\omega_0 \sim 2 \pi \times 9$ GHz and $J \sim 2 \pi \times 18$ MHz.
The total Hamiltonian is expressed in a rotating frame at the central frequency $\omega_0$, with the qubit excited-state ($|1\rangle$) energies positioned at $\omega_0 + \epsilon$. Due to the large difference in magnitude between $J$ and $\omega_0$, the counter-rotating terms can be neglected. The resulting Hamiltonian takes the form
\begin{align}
& H = \epsilon \left( \sigma^{+}_{Q_1} \sigma^{-}_{Q_1} + \sigma^{+}_{Q_2} \sigma^{-}_{Q_2} \right)  \notag \\
-& J \sum_{r,c < L} \Big( e^{i\phi_x(r,c)} a_{r,c+1}^{\dagger} a_{r,c}  
	+ e^{i\phi_y(r,c)} a_{r+1,c}^{\dagger} a_{r,c} + \text{h.c.} \Big) \notag \\
- & \sum_{\alpha = 1,2} \sum_{r,c} \left(
          g_{\alpha,r,c}     a_{r,c}           \sigma^{+}_{Q_{\alpha}} 
        + g_{\alpha,r,c}^{*} a_{r,c}^{\dagger} \sigma^{-}_{Q_{\alpha}} \right)\,. \label{eq:H_tot}
\end{align}
 The first term corresponds to bare energy of a qubit excitation. The second term represents the Hamiltonian of the isolated photonic lattice, where $a^{\dagger}$ and $a$ denote the photon creation and annihilation operators, respectively. The tunneling terms include Peierls phase factors \( \phi_x(r,c) \) and \( \phi_y(r,c) \), which encode synthetic gauge fields along the \( x \)- and \( y \)-directions indexed by row ($r$) and column ($c$). These phases are zero for tunneling between black sites, and take values in  \( \{0, \frac{\pi}{2}, \pi, \frac{3\pi}{2}\} \) for tunneling between blue and black neighboring sites. The final term captures the Jaynes–Cummings \cite{1443594} interaction between the qubits and the lattice, with $g_{\alpha,r,c}$ representing the coupling constants. Throughout this work, we focus on the single-excitation regime, where at most one excitation---either a qubit excitation or a photon---is present in the system at any given time. Unless stated otherwise, all energies are expressed in units of $J$.

The resonator coordinates are combined into a single site index $j(r, c) = (r-1)L + c$. Additionally, it is convenient to assign site indices to the qubits, denoted as $Q_1 = 0$ and $Q_2 = L^2 + 1$. We focus on odd lattice sizes \( L = 3, 5, 7, \dots \), for which the square lattice is invariant under rotations by angles $\pm \pi/2$ and $\pi$, such that the relative positions of the black and blue sites with respect to the lattice edges remain unchanged. This rotational symmetry will be used to simplify subsequent derivations. 

Analytical solutions for the Hofstadter lattice are known for rational magnetic flux values~\cite{Karnaukhov_2019, PhysRevLett.72.1890, PhysRevB.97.195439, PhysRevB.108.014204, doi:10.1143/JPSJ.79.124709}. For the configuration shown in Fig.~\ref{fig:qbt_hofst_lat}(a), corresponding to a flux of \(\frac{\pi}{2}\) and a large system size \(L \gg 1\), the bulk states form three well-separated energy bands (a detailed derivation is presented in Appendix \ref{ap_lat}):
\begin{subequations}
	\begin{align}
		0 \lesssim \left|\frac{E}{2J}\right| & \lesssim \sqrt{1 - \sqrt{\tfrac{1}{2}}}\,, \\
		\sqrt{1 + \sqrt{\tfrac{1}{2}}} & \lesssim  \left|\frac{E}{2J}\right| 
		\lesssim \sqrt{2}\,.
	\end{align}
\end{subequations}
Edge modes lie between these bands, as in Fig.~\ref{fig:qbt_hofst_lat}(b). The identical potential $\epsilon$ chosen for both qubits is tuned to lie near the center of the edge-mode region in the spectrum, thereby ensuring that the interaction predominantly involves edge states. Our analysis primarily focuses on the configuration where the qubits are coupled to opposite corners of the lattice, with equal coupling constants $g_{\alpha, 1, 1} = g_{\alpha, L, L} = g$. A detailed discussion of the more general case with unequal qubit potentials and coupling constants is provided in Appendix~\ref{ap1}.

To study energy transfer mediated by topological edge modes, we assume that each qubit interacts locally with a single photonic site located at the edge of the lattice, where chiral edge states are well localized.
At time $t = 0$, the system is initialized in a single-excitation state: qubit $Q_1$ is excited, qubit $Q_2$ is in the ground state, and the lattice is unoccupied.
In the absence of coupling $(g = 0)$, the two qubits form an isolated subsystem with degenerate energy level $\epsilon$; that is, both single-qubit excitation states have the same energy. When a weak coupling $g$ is introduced, this degeneracy is lifted through second-order virtual processes involving intermediate lattice excitations. These virtual photon exchange processes enable the excitation to propagate from $Q_1$ to $Q_2$ via the lattice. Since the total number of excitations is conserved, the dynamics can be described within the single-excitation subspace, leading to an effective Hamiltonian of the form:
\begin{align}
H_{\mathrm{eff}} = {} &\epsilon  \left( | Q_1 \rangle \langle Q_1| + | Q_2 \rangle \langle Q_2| \right) \nonumber \\
       {} & +  J_{\mathrm{eff}} \left( | Q_1 \rangle \langle Q_2| + | Q_2 \rangle \langle Q_1| \right)\,,
\end{align}upon integrating out the photons. Under this effective two-qubit Hamiltonian, the initial excitation is expected to undergo coherent oscillations between $Q_1$ and $Q_2$ with an effective frequency $\Omega_{\mathrm{eff}} = 2 J_{\mathrm{eff}}$. This ansatz neglects leakage outside of the $Q_1 - Q_2$ manifold, which we will discuss in more detail when treating the problem non-perturbatively.

\section{Perturbative limit}\label{sec_perturb}

For weak qubit-lattice coupling, \( g \ll J \), numerical time evolution reveals pronounced oscillations. Starting with \( \psi(Q_1, t = 0) = 1 \), the probability density primarily oscillates between the qubits, with minimal leakage into the lattice, as shown in Fig.~\ref{fig:oscilate}. These oscillations are stable when the qubit energy \( \epsilon \) is tuned near the center of the edge mode spectrum (Fig.~\ref{fig:qbt_hofst_lat}(b)), though the dynamics are highly sensitive to the precise value of \( \epsilon \).
\begin{figure}[b]
    \centering
	\includegraphics[width=0.95\linewidth]{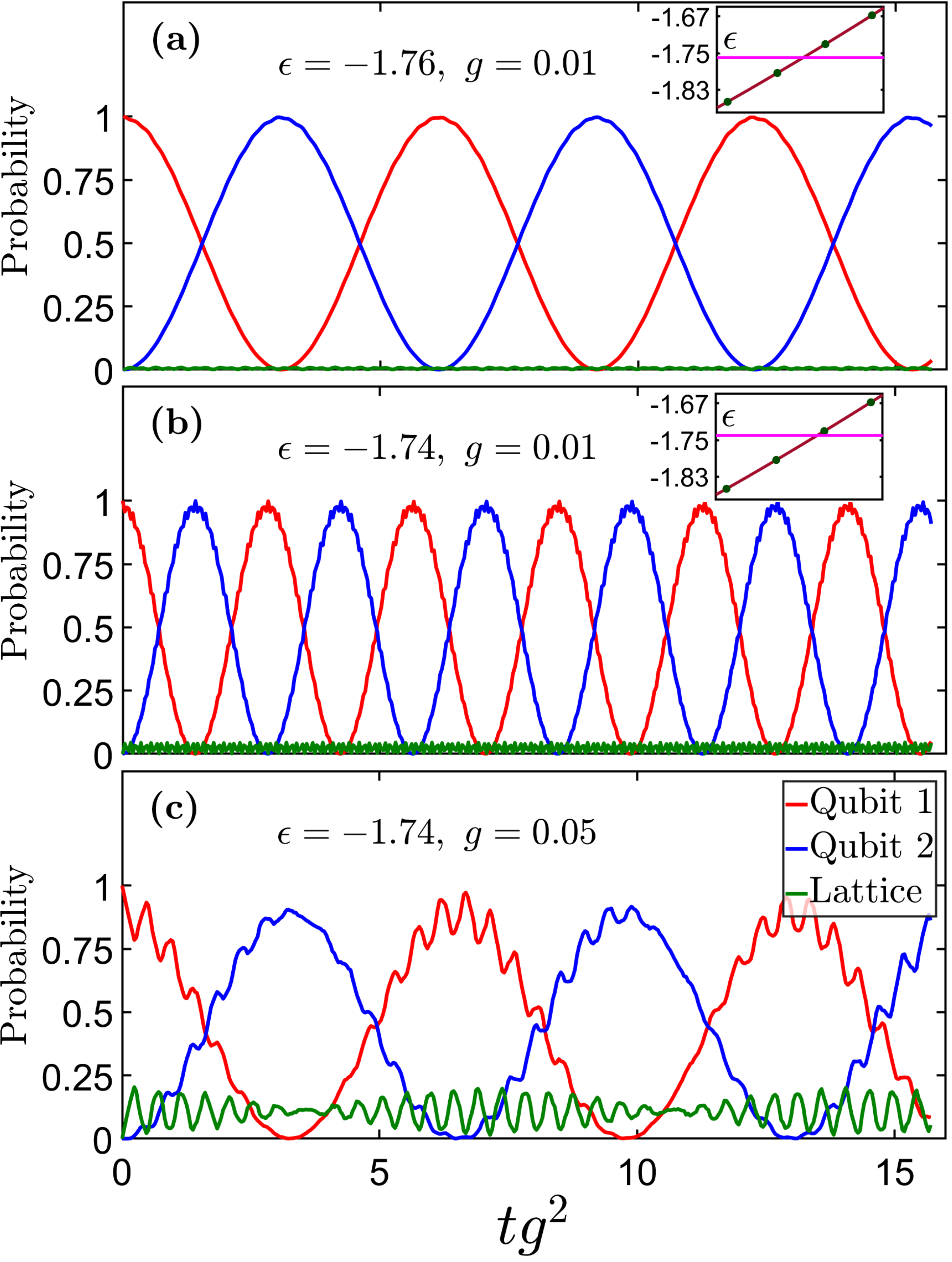}
	\caption{\label{fig:oscilate}Time evolution of qubit probabilities 
		\( P_{Q_1} = |\psi(Q_1,t)|^{2} \), 
		\( P_{Q_2} = |\psi(Q_2,t)|^{2} \), and total lattice population 
		\( P_\mathrm{lat} = \sum_{j}|\psi(j,t)|^{2} \) for \( L = 35 \) and \( J = 1 \). 
		(a) \( \epsilon = -1.76 \), \( g = 0.01 \); 
		(b) \( \epsilon = -1.74 \), \( g = 0.01 \); 
		(c) \( \epsilon = -1.76 \), \( g = 0.05 \). The insets in panels (a) and (b) display segments of the edge mode band structure, indicating the positions of the corresponding potentials $\epsilon$ (magenta line) relative to the lattice eigenvalues (green dots).}
\end{figure}
The unperturbed system hosts \( L^2 \) lattice eigenstates \( \psi_k \) and two degenerate qubit-localized states
\begin{align}\label{eq:corr_split1}
|\psi_{q\pm} \rangle = \frac{1}{\sqrt{2}} (|Q_1\rangle \pm |Q_2\rangle)\,, 
\end{align}with energy \( \epsilon \). The second-order correction to the degenerate level is:
{\small \begin{align} \label{eq:E_shift}
		\Delta^{(2)}_\pm(\epsilon) = \frac{g^2}{2} \Big( S_1 + S_2 \pm \sqrt{(S_1 - S_2)^2 + 4 |S_0|^2} \Big) \equiv g^2 f_\pm(\epsilon)\,,
\end{align}}with
\begin{subequations} \label{eq:abc-1}
	\begin{align}
		S_0(\epsilon) &= \sum_{n \ne q_\pm} \frac{\psi_n(1)\psi_n^*(L^2)}{\epsilon - E_n}\, ,\\
		S_1(\epsilon) &= \sum_{n \ne q_\pm} \frac{|\psi_n(1)|^2}{\epsilon - E_n} \, ,\\
		S_2(\epsilon) &= \sum_{n \ne q_\pm} \frac{|\psi_n(L^2)|^2}{\epsilon - E_n}\, .
	\end{align}
\end{subequations}

The system's symmetries can be exploited to simplify these results. The hopping terms in the lattice can be rotated around blue sites---e.g., by \(\frac{\pi}{2}\) counterclockwise as shown in Fig.~\ref{fig:qbt_hofst_lat}(a)---without modifying the system's physical properties. However, such a rotation introduces a phase shift of \(-\frac{\pi}{2}\) for any given eigenstate solution $\psi_n$, affecting only the blue sites; that is, \(\psi_n(b) \to \psi_n(b)e^{-i\frac{\pi}{2}}\). Performing this rotation twice at all blue sites is equivalent to a global rotation of the entire Hamiltonian matrix by \(\pi\), which swaps the indexes \(Q_1 \leftrightarrow Q_2\), \(1 \leftrightarrow L^2\), and so on, along with all the corresponding hopping terms. This transformation comes at the cost of flipping the sign of $\psi_n (j)$ values at the blue sites, i.e., \(\psi_n(b) \to -\psi_n(b)\). This implies a classification of the eigenstates into even (\(D_{+}\)) and odd (\(D_{-}\))  parity sectors. For all \(n \in D_{\pm}\), the eigenstates fulfill the \textcolor{black}{(generalized) parity} symmetry: 
\begin{align}\label{eq:glob_symm}
	\psi_{n}(L^{2}+1-j) =
	\begin{cases}
		\pm \psi_{n}(j) & \text{black sites, } Q_1, Q_2\, , \\
		\mp \psi_{n}(j) & \text{blue sites}\,.
	\end{cases}
\end{align}
A detailed proof is provided in Appendix~\ref{ap_sym}.

Owing to the equality in Eq.~(\ref{eq:glob_symm}), the functions \( f_\pm(\lambda) \) simplify as follows:
\begin{align}
	f_\pm(\lambda) = S_1(\lambda) \pm S_0(\lambda) 
	= 2 \sum_{n \in D_\pm} \frac{|\psi_n(1)|^2}{\lambda - E_n}\,.
\end{align}Since edge-mode eigenstates alternate in parity (for example $n\in D_+ \implies n+1 \in D_-$), and both \( \Delta E = E_{n+1} - E_n \) and \( |\psi_n(1)|^2 \) are approximately constant for states that are near the qubit energy, we can apply the identity \( \pi \cot(\pi z) = \frac{1}{z} + \sum_{n=1}^{\infty} \left( \frac{1}{z - n} + \frac{1}{z + n} \right) \) to approximate \( f_\pm(\lambda) \),  assuming (without loss of generality) that the nearest eigenvalue \( E_l < \lambda \) belongs to \( D_+ \):
\begin{subequations}
	\begin{align}
		f_+(\lambda) &\approx A_{\mathrm{f}} \tfrac{\pi}{J} \cot\left( \pi \tfrac{\lambda - E_l}{2\Delta E} \right) + \tfrac{B_{\mathrm{f}}}{J}\,, \label{eq:f_cot} \\
		f_-(\lambda) &\approx -A_{\mathrm{f}} \tfrac{\pi}{J} \tan\left( \pi \tfrac{\lambda - E_l}{2\Delta E} \right) + \tfrac{B_{\mathrm{f}}}{J}\,. \label{eq:f_tan}
	\end{align}
\end{subequations}where \( A_{\mathrm{f}} \approx \frac{J |\psi_k(1)|^2}{\Delta E} \sim 1 \) describes interactions mediated by virtual photon excitations and \( B_{\mathrm{f}} \) captures the Stark shift contribution  from distant states. These constants can be evaluated numerically, e.g., at \( \lambda^{\prime} = (E_l + E_{l+1})/2 \). Near \( \epsilon = -1.75J \), they converge (as \( L \gg 1 \)) to\begin{subequations}
\begin{align}
A_{\mathrm{f}} &= \frac{J}{{2 \pi}} \left( f_{+}(\lambda^{\prime}) - f_{-}(\lambda^{\prime}) \right) \approx 0.16 \,,  \\
B_{\mathrm{f}} &= \frac{J}{{2}} \left( f_{+}(\lambda^{\prime}) + f_{-}(\lambda^{\prime}) \right) \approx -0.31 \,.
\end{align}\end{subequations}

Another relevant quantity, converging for large system sizes, is the density of edge states in units of $J/L$:
\begin{align} \label{eq:def_D_f_def}
	\rho_{\mathrm{e}} \equiv \frac{J}{L \cdot \Delta E } \approx 0.46 
	\quad \text{(near \( \epsilon = -1.75J,\ L \gg 1 \))}\, .
\end{align}

The oscillations in Fig.~\ref{fig:oscilate} are primarily driven by the two qubit states defined in Eq.~(\ref{eq:corr_split1}). The corresponding effective oscillation frequency is:
\begin{align}
    \label{Omg_eff_pert}
	\Omega_{\mathrm{eff}}(g,\epsilon) 
	= g^2 |f_+(\epsilon) - f_-(\epsilon)| 
	\approx \frac{2\pi A_{\mathrm{f}}\ \textcolor{black}{g^2} }{J \sin\left( \pi \tfrac{\epsilon - E_l}{\Delta E} \right)}\,.
\end{align}

\textcolor{black}{This aligns with the earlier observation that the oscillation frequency is highly sensitive to the precise placement of the potential $\epsilon$ relative to the lattice eigenvalues.
For clarity, we define a \textit{midpoint} as the energy value halfway between two consecutive lattice eigenvalues $(E_{l} + E_{l+1})/2$, and a \textit{resonance} as a case where $\epsilon$ coincides exactly with one of the eigenvalues $\epsilon = E_{l}$. As shown in Eq.~\eqref{Omg_eff_pert}, the effective frequency $\Omega_{\mathrm{eff}}$ reaches its minimum when $\epsilon$ is placed at the midpoint, corresponding to $(\epsilon - E_l)/\Delta E = 1/2$.}

At resonance (\( \epsilon = E_l \)), a three-fold degeneracy arises and is lifted at first order as \( \Delta^{(1)} = 0, \pm \Omega_0 \), with \( \Omega_0 = g |\psi_l(1)| \sqrt{2} \). This produces a richer three-state oscillation pattern, as shown in Fig.~\ref{fig:Rsn_oscil}:
\begin{subequations}
	\begin{align}
		|\Psi(Q_1)|^2 &= \cos^4\left( \tfrac{\Omega_0 t}{2} \right)\,, \\
		|\Psi(Q_2)|^2 &= \sin^4\left( \tfrac{\Omega_0 t}{2} \right)\,, \\
		\sum_{\text{lattice}} |\Psi(j)|^2 &= \tfrac{1}{2} \sin^2(\Omega_0 t)\,.
	\end{align}
\end{subequations}
In this regime, a significant portion of the wavefunction amplitude periodically spreads into the main lattice.\begin{figure}[t]
\centering
	\includegraphics[width=0.95\linewidth]{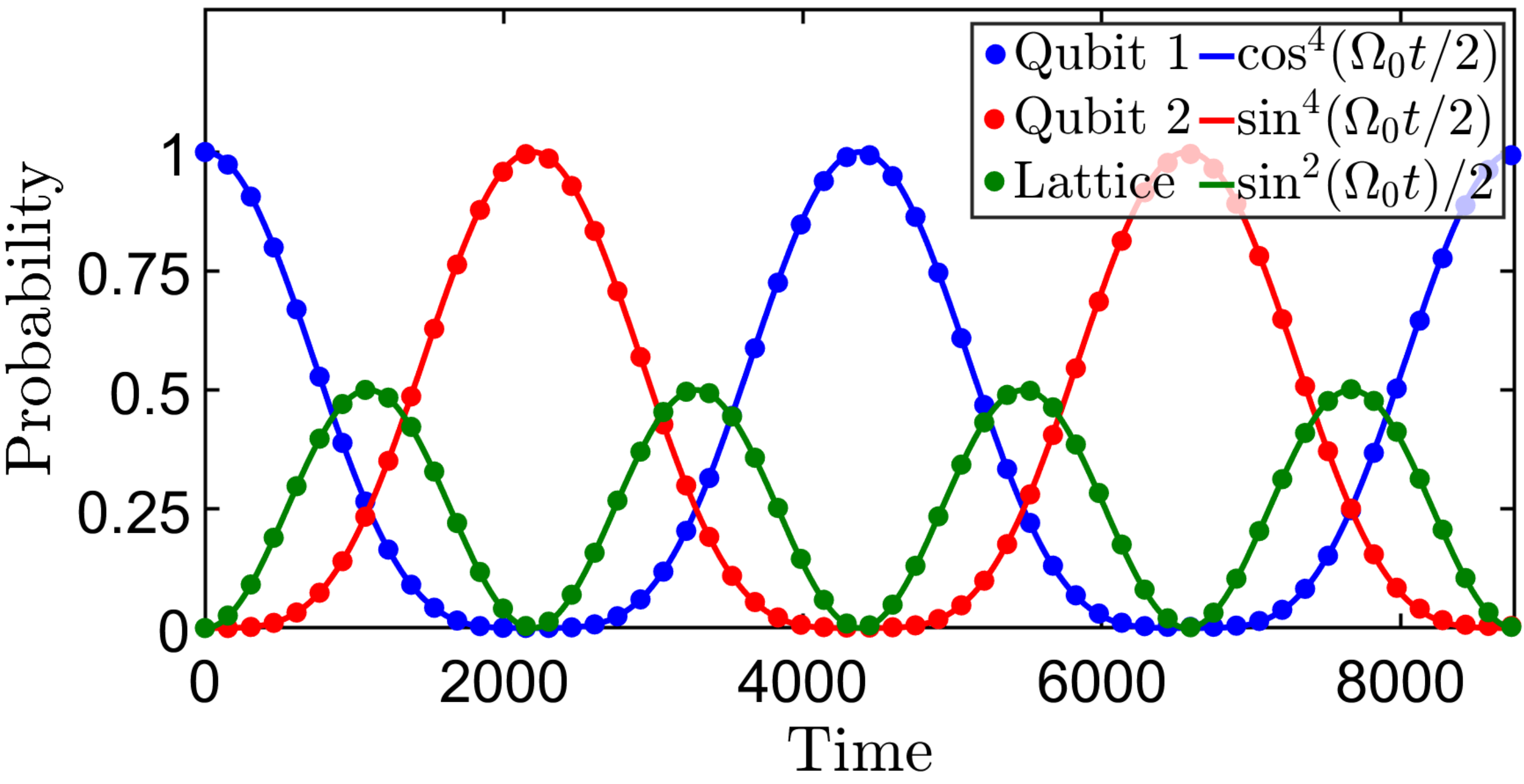}
\caption{\label{fig:Rsn_oscil}
Oscillations induced when the qubits are brought into resonance with an edge mode at $\epsilon=E_{l}=-1.7307$, with a coupling strength $\Omega_0 = g\sqrt{2}|\psi_{l}(1)|$, and an oscillation period $T=2\pi/ \Omega_0 \approx 4382$. The system size is $L=35$, and the coupling constant is $g=0.01$.}
\end{figure}

\section{Non-perturbative solution}\label{sec_finite}

\textcolor{black}{
Experimentally, one can readily access the non-perturbative regime. Figs.~\ref{fig:oscilate}(c) and ~\ref{fig:Rsn_oscil} illustrate examples of dynamics that go beyond the simple perturbative scheme based on a two-fold degeneracy lifted at second order.} A more general solution is required---one that can accurately describe the system’s behavior across all regimes. Persisting with the higher orders of perturbation theory leads to increasingly cumbersome calculations without yielding a definitive or comprehensive understanding of the underlying dynamics. Instead, we adopted a radically different approach to go beyond the limitations of perturbation theory and derive an analytical solution applicable to finite and even large values of $g$.

We decompose the system of \( L^{2}+2 \) linear equations \( \hat{H}|\psi\rangle = \lambda|\psi\rangle \) into two groups: \( L^{2}-2 \) equations that are independent of \( \epsilon \) and \( g \) (for the inner lattice sites), and four equations involving \( \lambda \psi(1), \lambda \psi(L^{2}), \lambda \psi(Q_{1}), \lambda \psi(Q_{2}) \) on the right-hand side. The first group can be solved for any given \( \lambda \), satisfying one of the \textcolor{black}{parity} symmetry conditions from Eq.~(\ref{eq:glob_symm}). For both the odd and even branches of solutions \( \psi_{\lambda \pm} \), any relationship between the main lattice sites depends solely on \( \lambda \):
\begin{equation}
	\frac{\psi_{\lambda\pm}(i)}{\psi_{\lambda\pm}(j)} = \text{func.}(\lambda), \ \text{not} \ (\epsilon, g)\,.\label{eq:relation_lambda}
\end{equation}
The remaining equations act as boundary conditions that restrict the choices of \( \lambda \):\begin{figure}[t]
	\centering
	\includegraphics[width=0.95\linewidth]{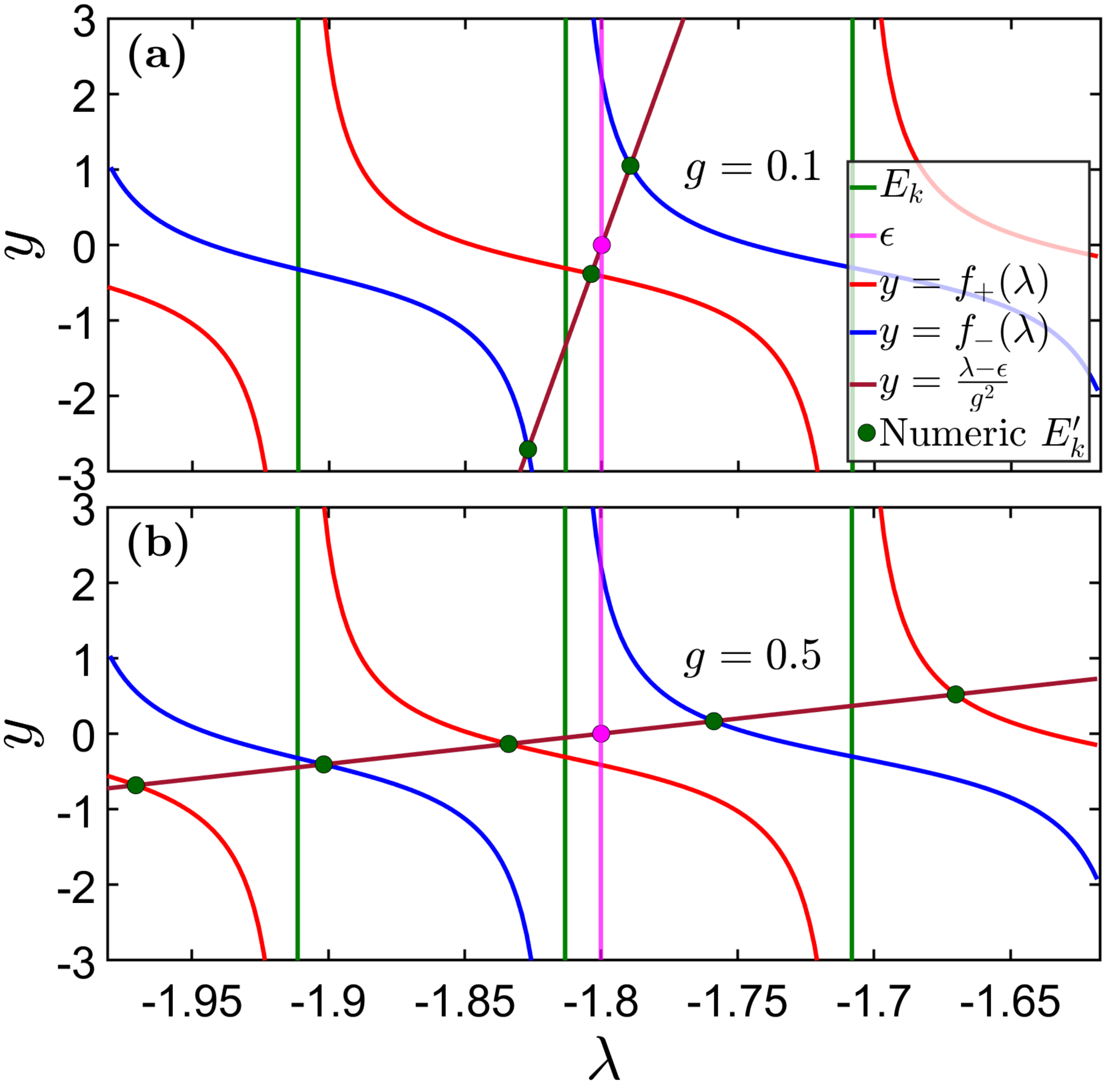}
	\caption{\label{fig:y_g_f}Numerical results for eigenvalues $E^{\prime}_k$ compared with the relation in Eq.~(\ref{eq:bc_4}). It demonstrates exact agreement with the theoretical prediction, with each solution for $E^{\prime}_k$  lying at the intersection of the functions of $y=f_{\pm}(\lambda)$ and the line $y=\frac{\lambda - \epsilon}{g^2}$.  \( f_{\pm}(\lambda) \) is calculated numerically in the deep perturbative limit $g \to 0$, as in Eq.~(\ref{eq:E_shift}). $E_k$ are the unperturbed eigenvalues of the lattice. The lattice size is \( L = 21 \), and the qubit potential is \( \epsilon = -1.8 \). Solutions work equally well for weak coupling (a) \( g = 0.1 \) or moderate/strong coupling (b) \( g = 0.5 \).}
\end{figure}

\begin{subequations}\begin{align}
\epsilon \psi(Q_{1}) - g \psi(1) = {} &\lambda \psi(Q_{1})\,,
\label{eq:Schrod_1_a}
\\
-J \psi(2) - J \psi(1+L) - g \psi(Q_{1}) = {} &\lambda \psi(1) \,,
\label{eq:Schrod_1_b}
\end{align}
\end{subequations}
leading to\begin{equation}
\frac{g^{2}}{\lambda - \epsilon} - \lambda =
\textcolor{black}{J}
\frac{\psi(2) + \psi(1+L)}{\psi(1)}\,.\label{eq:bc_1}
\end{equation}
Substituting the solution for infinitesimally small \( g \) from Eq.~(\ref{eq:E_shift}) into Eq.~(\ref{eq:bc_1}), and treating $f_{\pm}(\lambda)$ as a known function, yields an equation for all possible eigenvalues \( \lambda \):
\begin{equation}
	 f_{\pm}(\lambda) = \frac{\lambda - \epsilon}{g^{2}}\,.\label{eq:bc_4}
\end{equation}
Due to the \textcolor{black}{parity} symmetry in Eq.~\eqref{eq:glob_symm}, Eqs.~\eqref{eq:Schrod_1_a} and ~\eqref{eq:Schrod_1_b} written for the second qubit will give the same expression in \eqref{eq:bc_4}.
The solution of \ref{eq:bc_4} is graphically illustrated in Fig.~\ref{fig:y_g_f}.

In theory, Eq. \eqref{eq:bc_4} provides the solution for all eigenstates contributing to the superposition and is valid for arbitrary input parameters, assuming the explicit form of $f_{\pm}(\lambda)$ is known. However, in this work, we focus on configurations predominantly governed by edge states, using Eqs. \eqref{eq:f_cot} and \eqref{eq:f_tan} to approximate $f_{\pm}(\lambda)$. Treating the terms associated with $B_{\mathrm{f}}$ as a constant energy offset, we define the adjusted potential
\begin{equation}
\tilde{\epsilon} = \epsilon + B_{\mathrm{f}} \frac{g^2}{J}    
\end{equation}
and introduce the following dimensionless variables:
\begin{equation}
w = \frac{\tilde{\epsilon} - E_l}{\Delta E}\,, \ \ \ \, 
x = \frac{\lambda - E_l}{\Delta E}\,, \ \ \ \, 
G_0 = \frac{g^2 L}{J^2} \label{eq:gLJ}\, .
\end{equation}
Rewriting Eq. \eqref{eq:bc_4} in terms of these variables yields an equation that depends only on the dimensionless qubit-photon coupling $G_0$:
\begin{equation}
x - w = G_0 \pi A_{\mathrm{f}} \rho_{\mathrm{e}} \cot\left(\frac{\pi}{2} (x-n)\right)  \,,\label{eq:scale_eq1}
\end{equation}
where $n=0,\pm 1, \pm2, ...$ indexes all consecutive branches of the functions $f_{\pm}(\lambda)$.

It can be observed that Eq.~\eqref{eq:scale_eq1} yields two distinct solutions for $x$, associated with the energies $\lambda_{-}$ and $\lambda_{+}$, which lie immediately below and above the adjusted potential $\tilde{\epsilon}$, respectively. In the perturbative regime, this pair of energies lies closest to the original degenerate level $\epsilon$; however, as $g$ increases, the pair shifts away from the bare qubit potential by $B_\mathrm{f}g^2/J$. Both $\lambda_{\pm}$  fall within the same energy interval $\Delta E$ of the lattice eigenvalues, such that $E_l \le \lambda_{-} \le \tilde{\epsilon} \le \lambda_{+} \le E_{l+1}$. It is therefore natural to define the dominant oscillation frequency at finite $g$ as the eigenvalue separation: $\Omega_{\mathrm{eff}} =  \lambda_{+} - \lambda_{-}$. As will be shown later, $\lambda_{\pm}$ are not merely mathematically convenient---they correspond to the two eigenstates with the highest qubit population.

Applying the scaling technique outlined in Eq. \eqref{eq:scale_eq1}, $\lambda_{\pm}$ translate into $x_n$ solutions for $n = 0$ and $n = 1$, while $\Omega_{\mathrm{eff}}$ scales as $\frac{\Omega_{\mathrm{eff}}}{\Delta E} = x_0 - x_1$ leading to:
\begin{align}\label{eq:slop1}
	G_0 =\frac{
        \frac{\Omega_{\mathrm{eff}}}{\Delta E}
        \left[\sin\left( \frac{\pi}{2}(x_0 + x_1)  \right) + \sin\left( \frac{\pi}{2}  \frac{\Omega_{\mathrm{eff}}}{\Delta E} \right)\right] }{ 2 \pi \rho_{\mathrm{e}} A_{\mathrm{f}} \cos\left( \frac{\pi}{2} \cdot \frac{\Omega_{\mathrm{eff}}}{\Delta E} \right) }. 
\end{align}
For a fixed $(x_0 - x_1)$, the term $\sin\left( \frac{\pi}{2} (x_0 + x_1) \right)$ reaches its maximum value of $1$ when the adjusted potential lies exactly at the midpoint between the lattice eigenstates, i.e., $\tilde{\epsilon} = E_l + \Delta E (\frac{x_0 + x_1}{2}) = \frac{E_{l+1} + E_{l}}{2}$. On the other hand, it attains its minimum value $\sin\left( \frac{\pi}{2}\frac{\Omega_{\mathrm{eff}}}{\Delta E} \right)$ in the resonant case, when one eigenvalue lies at the boundary of the energy interval, $x_0 + x_1 = \frac{\Omega_{\mathrm{eff}}}{\Delta E} + 2k$, where $k = 0,1$. Therefore, the minimum and maximum values of the parameter $G_0 =  g^{2}L/J^{2}$ for a given frequency can be expressed as:
\begin{subequations}
	\begin{align}
	\frac{g_{\mathrm{min}}^{2}L}{J^{2}} &= \frac{1}{\pi A_{\mathrm{f}} \rho_{\mathrm{e}}} \left( \frac{\Omega_{\mathrm{eff}}}{\Delta E} \right) \tan\left( \frac{\pi}{2} \cdot \frac{\Omega_{\mathrm{eff}}}{\Delta E} \right), \label{eq:g_omg_min} \\
	\frac{g_{\mathrm{max}}^{2}L}{J^{2}} &= \frac{1}{\pi A_{\mathrm{f}} \rho_{\mathrm{e}}} \left( \frac{\Omega_{\mathrm{eff}}}{\Delta E} \right)
	\cdot \frac{ 1 + \sin\left( \frac{\pi}{2} \cdot \frac{\Omega_{\mathrm{eff}}}{\Delta E} \right) }{ 2 \cos\left( \frac{\pi}{2} \cdot \frac{\Omega_{\mathrm{eff}}}{\Delta E} \right) }. \label{eq:g_omg_max}
\end{align}\end{subequations}
These expressions are shown to match the exact numerical simulations in Fig. 5.
Notably, there is a significant difference between the minimum and maximum values of the effective frequencies $\Omega_\mathrm{eff}$ for small values of the coupling parameter $G_0$, consistent with the perturbative regime. In this limit, the second-order energy correction $\Delta(\epsilon) \propto g^2$ transitions to a first-order correction $\Delta(\epsilon) \propto g$ as the configuration approaches resonance. As $G_0$ increases, the effective frequency becomes less sensitive to the exact placement of the potential $\epsilon$ relative to the lattice eigenvalues and approaches its upper bound, with $\frac{\Omega_{\mathrm{eff}}}{\Delta E} \to 1$.
\begin{figure}[t]
		\centering
	\includegraphics[width=0.95\linewidth]{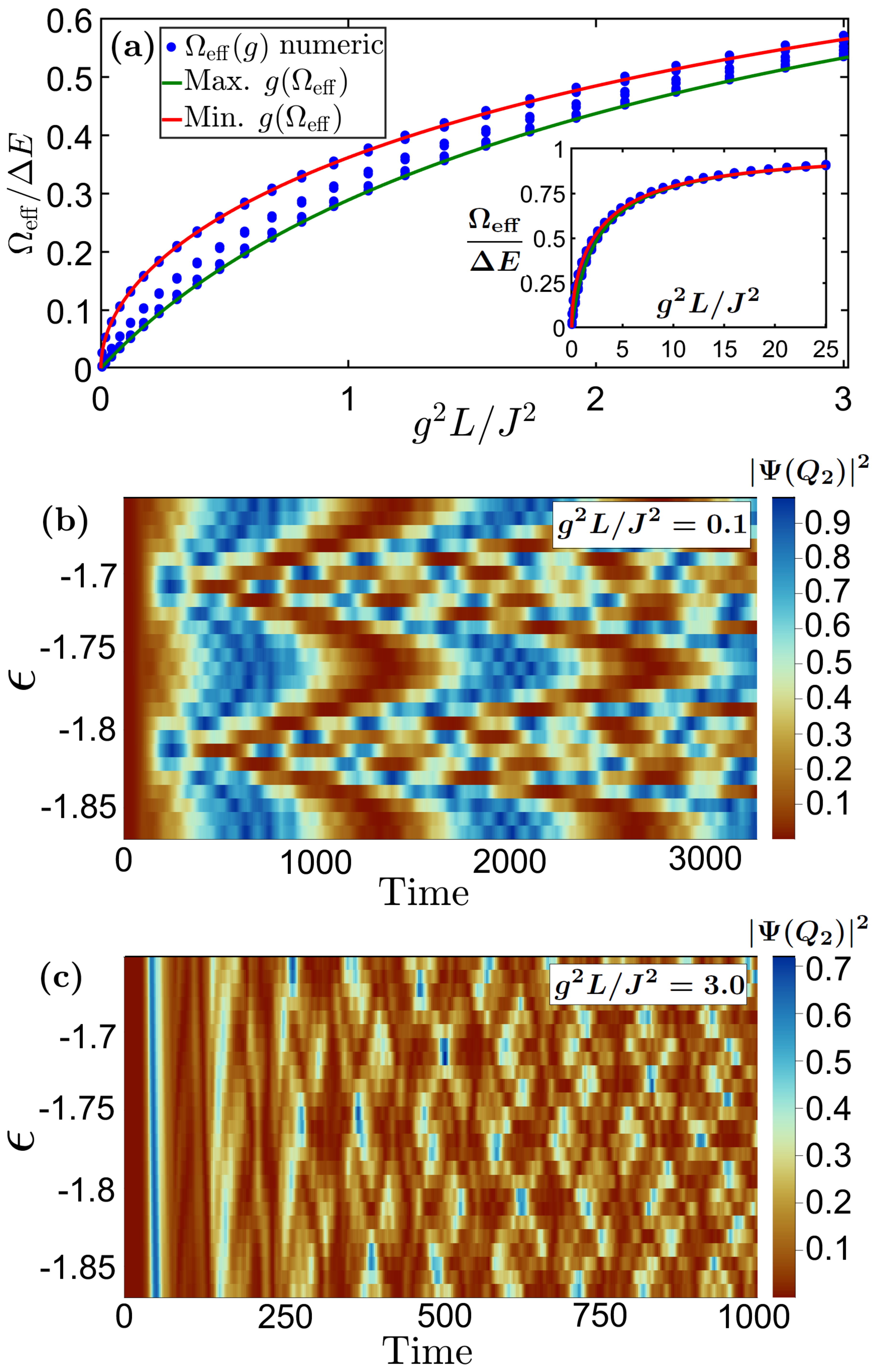}
	\caption{\label{fig:Omg_eff_big_g}
		(a) Comparison of numerical and analytical bounds, Eqs. (\ref{eq:g_omg_min}) and (\ref{eq:g_omg_max}), for the effective frequency $\Omega_{\mathrm{eff}}$ at given $g^{2}L/J^2$, calculated for $L=21$ between $E_l = -1.81$ and $E_{l+1} = -1.71$. The analytical predictions use $\tilde{\epsilon} = E_l$ (resonance) for minimum $g(\Omega_{\mathrm{eff}})$ and $\tilde{\epsilon} = (E_l + E_{l+1})/2$ (midpoint) for maximum $g(\Omega_{\mathrm{eff}})$. The inset shows the same comparison over a wider range of $g^{2}L/J^2$ values.  
		(b,c) Time evolution of the qubit population $|\Psi(Q_2, t)|^2$ for various $\epsilon$, shown for $g^{2}L/J^2 = 0.1$ and $3.0$. In all cases, $J=1$. 
	}
\end{figure}

\section{Fidelity}\label{sec_fidelity}
An important characteristic of the induced interactions is the fraction of probability that remains localized on the qubits throughout the oscillation, as opposed to the portion lost to the lattice, which we refer to as the oscillation fidelity. Ideally, the fidelity should achieve a maximum value of $1$ in the non-resonant $g\to 0$ limit (Fig.~\ref{fig:oscilate}(a)), and progressively decrease for finite $g$ values (Fig.~\ref{fig:oscilate}(c)) or under parameter combinations that induce resonance (Fig.~\ref{fig:Rsn_oscil}). To quantify this behavior, we consider the time-averaged probability, for remaining within the qubit manifold:
\begin{equation}
\textcolor{black}{
    F = \lim_{T\ \to \infty} \frac{1}{T}\int_{0}^{T}
	\left( |\Psi(Q_1, t)|^2 + |\Psi(Q_2, t)|^2 \right) dt. \label{eq:F_average}}
\end{equation}This quantity is equivalent to the oscillation fidelity, as the occupations of each qubit show symmetric oscillations. Building on the same approach that led to Eq. \eqref{eq:bc_4}, we derive explicit expressions for the qubit site occupations. In brief, we employ infinitesimally small perturbations---not to develop a full perturbative solution, but rather as a tool to obtain the relationships between different lattice sites for a given $\lambda$.

For the infinitesimal first-order perturbation of the states \( \psi_{q_{\pm}} \) (Eq.~(\ref{eq:corr_split1})), we have:
\begin{equation}
	|\psi_{q_{\pm}}^{\prime}\rangle = |\psi_{q_{\pm}}\rangle + \sum_{n \ne q_{\pm}} |\psi_n\rangle \frac{g \sqrt{2} \psi_n(1)}{\epsilon - E_n} + O(g^{2})\,.
\end{equation}
The total density transferred from the qubit sites to the lattice for state \( \psi_{q_{\pm}} \), resulting from the perturbation at a given \( \epsilon \), can be calculated to leading order \( O(g^{2}) \) as:
\begin{equation}
	\Delta P_{\text{lat}}(\epsilon) = \sum_{n \in D_{\pm}} \frac{2g^{2} |\psi_n(1)|^2}{(\epsilon - E_n)^2} = -g^{2} \frac{\partial f_{\pm}(\epsilon)}{\partial \epsilon}\,.
\end{equation}
Additionally, by applying the Eq.~(\ref{eq:Schrod_1_a}) in the limit \( g \to 0 \), we find the total density transferred to the first site as:
\begin{equation}
	\Delta|\psi_{q_{\pm}}(1)|^2 = g^{2} f_{\pm}^2(\lambda) |\psi_{q_{\pm}}(Q_1)|^2 = \frac{g^{2}}{2} f_{\pm}^2(\epsilon)\,.
\end{equation}
According to Eq.~(\ref{eq:relation_lambda}), this relationship between the probability at the first site and the total probability across the entire lattice holds for any eigenstate characterized by a given \( \lambda \) and a specified $\pm$ parity:
\begin{equation}
	\frac{|\psi_{\lambda}(1)|^2}{P_{\text{lat}}(\lambda)} = -\frac{1}{2} \frac{f_{\pm}^2(\lambda)}{\frac{\partial f_{\pm}(\epsilon)}{\partial \epsilon}}\,.\label{eq:prob_relate}
\end{equation}
Finally, using Eq.~(\ref{eq:Schrod_1_a}) for finite \( g \), together with Eq. \eqref{eq:prob_relate} and the normalization condition \( 2|\psi_{\lambda}(Q_1)|^2 + P_{\text{lat}}(\lambda) = 1 \), we can derive the occupation of each qubit site for any given eigenstate \( \psi_{\lambda} \):
\begin{equation}
|\psi_{\lambda}(Q_1)|^2 = \frac{1}{2 \left( 1 - g^2 \frac{\partial f_{\pm}(\lambda)}{\partial \lambda} \right)}\,.\label{eq:psi_Q_1_2}
\end{equation}
Treating each solution \( \lambda \) in Eq. \eqref{eq:bc_4} as a continuous function of \( g \) and \( \epsilon \) (\( \lambda = \lambda(g, \epsilon) \)), we can write relations for partial derivatives that are convenient for this analysis:
\begin{subequations}\begin{align}
		\mathrm{d}\lambda - \mathrm{d}\epsilon = {} &g^{2} \frac{\partial f_{\pm}(\lambda)}{\partial \lambda} \mathrm{d}\lambda + f_{\pm}(\lambda) \mathrm{d}(g^{2})\, ,\\
		\frac{\partial \lambda}{\partial \epsilon} = {} &\frac{1}{1 - g^{2} \frac{\partial f_{\pm}(\lambda)}{\partial \lambda}}\, , \label{eq:par_epsilon}\\
		\frac{\partial \lambda}{\partial (g^{2})} = {} &\frac{f_{\pm}(\lambda)}{1 - g^{2} \frac{\partial f_{\pm}(\lambda)}{\partial \lambda}}\,.
	\end{align}
\end{subequations}
By substituting Eq. \eqref{eq:par_epsilon} into Eq. \eqref{eq:psi_Q_1_2}, and employing the scaled variables defined in Eq.~\eqref{eq:gLJ}, the expressions for the qubit probabilities can be simplified to the form:
\begin{equation}
|\psi_{\lambda}(Q_1)|^2 = \frac{1}{2} \frac{\partial \lambda}{\partial \epsilon} = \frac{1}{2} \frac{\partial x}{\partial w}\,.\label{eq:scale_eq2}
\end{equation}
The sum of all \( |\psi_{\lambda}(Q_1)|^2 \) must be equal to 1, which is evident since \( \sum_{n} \lambda_n = 2\epsilon \).

To evaluate the fidelity $F$, as defined in Eq.~\eqref{eq:F_average}, we analyze the time-dependent wavefunction $\Psi(j,t)$ which is expressed as a superposition of the system’s eigenstates $\psi_n(j)$:
\begin{equation}
	\Psi(j,t) = \sum_n c_n \psi_n(j) e^{-i\lambda_n t}\,,
\end{equation}
with the initial condition at $t=0$ corresponding to full population of the first qubit, i.e., $\sum_n c_n \psi_n(j) = \delta_{j,Q_1}$. Assuming that the phases $e^{-i\lambda_n t}$ are non-degenerate, we can average them at long times and use the \textcolor{black}{parity} symmetry property in Eq.~(\ref{eq:glob_symm}) to obtain
\begin{align}
	F = 2 \sum_n |c_n|^2 |\psi_n(Q_1)|^2 = \frac{1}{2} \sum_n \left(\frac{\partial\lambda_n}{\partial\epsilon}\right)^2. \label{eq:F_expr}
\end{align}
In the perturbative regime $g \to 0$, we observe $F \to 1$ for non-resonant cases, while under resonance conditions the fidelity approaches $F \to \frac{3}{4}$.\begin{figure}[b]
	\centering
	\includegraphics[width=0.95\linewidth]{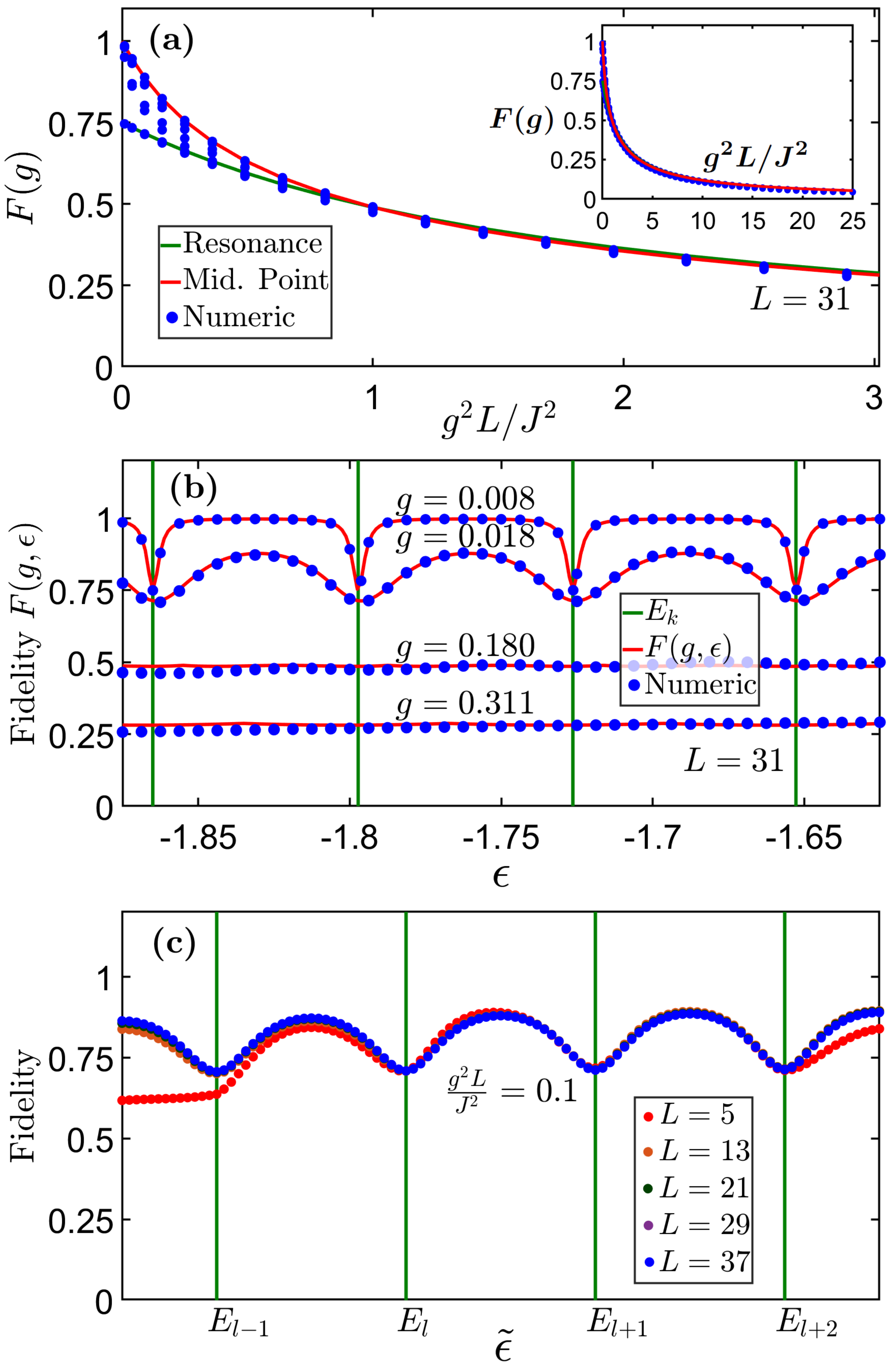}
	\caption{\label{fig:fidelity_comb}
		(a) Comparison of the fidelity range obtained numerically with the analytical approximation in Eq.~(\ref{eq:F_expr}) for various values of $g^{2}L/J^2$. Numerical data are generated for 6 evenly spaced values of the adjusted potential $\tilde{\epsilon}$ between two consecutive eigenenergies of the isolated lattice ($E_l = -1.80$, $E_{l+1} = -1.73$) with $L=31$. The analytical curves are based on $\tilde{\epsilon}=E_l$ (resonance) and $\tilde{\epsilon}=(E_l + E_{l+1})/2$ (midpoint), capturing the extrema of fidelity up to $g^{2}L/J^2 \approx 1$. The inset shows the same comparison over a wider range of $g^{2}L/J^2$ values.
		(b) Fidelity values obtained numerically for various combinations of $\epsilon$ and $g$, compared with corresponding analytical predictions for the same lattice size $L=31$. 
		(c) Numerical fidelity results for lattice sizes $5 \leq L \leq 37$, keeping $g^{2}L/J^2 = 0.1$ constant. The $\tilde{\epsilon}$ ranges are rescaled to span the same number of eigenstates for each $L$, maintaining the central part of the spectrum around $E_l < -1.75 < E_{l+1}$. This panel demonstrates the rapid convergence of fidelity as $L$ increases. All panels use $J=1$.}
\end{figure}

To evaluate fidelity beyond the perturbative regime, we consider an approximate analytical treatment. Each eigenvalue $\lambda_n (x_n)$
within the edge mode spectrum can be determined using Eq. \eqref{eq:scale_eq1} by applying the approximation:
\begin{align}
    \cot\left(\frac{\pi}{2}x\right) \approx \frac{2}{\pi} \left( \frac{1}{x} - x \right),\ \  \mbox{with}\ \  0 \le x \le 1\,.
    \label{eq:cot_aprxm}
\end{align}
The choice of Eq. \eqref{eq:cot_aprxm} was motivated by its close agreement with the target function in both value and derivative in the region where $\cot(\pi x / 2)$ is large. Additionally, it matches the target function at both endpoints of the interval $[0, 1]$, which is particularly important for accurately identifying solution intersections. The maximum deviation within this interval does not exceed $0.048$.

Substituting Eq.~(\ref{eq:cot_aprxm}) into Eq.~(\ref{eq:scale_eq1}), we obtain:
\begin{align}
	x - w &= C_0\left( \frac{1}{x - n} - (x - n) \right),
\end{align}
where $C_0 = 2 G_0 A_{\mathrm{f}} \rho_{\mathrm{e}}$. This yields two families of solutions depending on their location with respect to $w$:
{\small \begin{subequations}
	\begin{align}
	x_b(n) &= n + \frac{(w - n) - \sqrt{(w - n)^2 + 4C_0(C_0 + 1)}}{2(C_0 + 1)}, \quad n \le 1, \label{eq:x_b} \\
	x_a(n) &= n + \frac{(w - n) + \sqrt{(w - n)^2 + 4C_0(C_0 + 1)}}{2(C_0 + 1)}, \quad n \ge 0. \label{eq:x_a}
\end{align}\end{subequations}}Assuming negligible contributions from states outside the edge-mode spectrum (i.e., $|\psi_n(Q_1)|^2 \approx 0$ for such states), the normalization condition $\frac{1}{2} \sum_n \frac{\partial x_n}{\partial w} = 1$ must hold, resulting in
\begin{align}\label{eq:sum_2}
	&\sum_{n \le 1} \left( 1 - \frac{w - n}{\sqrt{(w - n)^2 + 4C_0(C_0 + 1)}} \right) \nonumber \\  
  + & \sum_{n \ge 0} \left( 1 + \frac{w - n}{\sqrt{(w - n)^2 + 4C_0(C_0 + 1)}} \right) \approx 4(C_0 + 1). 
\end{align}
Then, considering Eqs.~(\ref{eq:sum_2}) and (\ref{eq:scale_eq2}), and the known identity \( \sum_{n=-\infty}^{\infty}\frac{1}{(n - z)^{2} + r^{2}} = \frac{\pi \sinh(2\pi r)}{r\left[\cosh(2\pi r) - \cos(2\pi z)\right]} \ \ \mathrm{with}\ r \ge 0 \), the fidelity function $F(g,\epsilon)$ can be approximated as:
\begin{align}
	F &= \frac{1}{2} \sum_{x_{a}} \left( \frac{\partial x_{a}}{\partial w} \right)^2 + \frac{1}{2} \sum_{x_{b}} \left( \frac{\partial x_{b}}{\partial w} \right)^2 \nonumber \\
	&\approx \frac{1}{C_0 + 1} \left[ 1 - \frac{C_0}{2} \left( 
	\frac{\pi}{R_0} \frac{\sinh(2\pi R_0)}{\cosh(2\pi R_0) - \cos(2\pi w)} \right. \right. \nonumber \\
	&\qquad \left. \left. + \frac{1}{w^{2} + R_0^{2}} + \frac{1}{(w - 1)^{2} + R_0^{2}} \right) \right], \label{eq:fidel_anlt}
\end{align}
with $R_0 = \sqrt{{4 C_0}(C_0 + 1)}$. This formulation offers a scalable analytical approximation for the fidelity, valid across a wide range of coupling strengths and lattice sizes, and it closely matches the numerical results presented in Fig.~\ref{fig:fidelity_comb}.

We observe a significant difference in fidelity at small values of the parameter $ G_0 = g^2 L / J^2$, consistent with perturbative results, which show a sharp drop in fidelity from 1 to 0.75 in the resonance case. As $G_0$ increases, the fidelity becomes less sensitive to the exact position of $\epsilon$, in agreement with the behavior of the effective frequency shown in Fig.~\ref{fig:Omg_eff_big_g}. Around $G_0 \approx 1$, the fidelity becomes nearly independent of $\epsilon$ and continues to decrease steadily as $G_0$ increases.
Fig. \ref{fig:fidelity_comb}(c) demonstrates the rapid convergence of fidelity function for $L \gg 1$, confirming the validity of our approximation for large lattice sizes.

Notably, substituting Eqs. \eqref{eq:x_b} and \eqref{eq:x_a} directly into Eq. \eqref{eq:scale_eq2} supports our earlier assumption that the solutions $x_b(1)$ and $x_a(0)$, which lie closest to the adjusted potential $w(\tilde{\epsilon})$,  correspond to the eigenstates with the highest qubit population.

\textcolor{black}{
The connection between the loss of fidelity and increasing values of $\Omega_{\mathrm{eff}} / \Delta E$ can also be understood in terms of the finite propagation time of edge excitations.
Specifically, information cannot travel between qubits faster than the group velocity $v_g = \frac{\partial E}{\partial k_x}$ of the edge modes.
From Fig.~\ref{fig:qbt_hofst_lat}(a), we extract a group velocity of $v_g \approx 0.63 J$ for the targeted $\epsilon = -1.75 J$, where length is measured in units of the two-site unit cell.
This implies that the minimum time required for an excitation to travel a distance of $2L$, from one corner of the lattice to the opposite corner, is approximately $L/v_g$.
To maintain coherent transfer, this propagation time should be safely less than half the oscillation period, that is, $L/v_g < \pi / \Omega_{\mathrm{eff}}$.
This condition imposes a constraint on the effective frequency: $\Omega_{\mathrm{eff}} / \Delta E < \pi \rho_{\mathrm{e}} v_g / J \approx 0.92,$
which is consistent with the degradation of fidelity observed for higher values of $\Omega_{\mathrm{eff}} / \Delta E$, which correspond to larger values of $g^2 L / J^2$,
as evident from Figs.~\ref{fig:fidelity_comb}(a) and \ref{fig:Omg_eff_big_g}(a). More generally, the scaling relation $\Omega_\mathrm{eff} \propto \Delta E \propto 1/L$ is precisely the scaling imposed by a finite group velocity, as required by relativistic (or Lieb-Robinson) bounds.
}

In fact, most of the results discussed above are largely insensitive to the spatial separation along the edge between the two ``connection'' sites where the qubits couple to the lattice. To demonstrate this, consider a modified configuration in which the qubits are connected symmetrically to a pair of sites separated by a distance $d$ (in units of the lattice spacing) along one of the longer edges of the rectangular lattice, as illustrated in Fig.~\ref{fig:depend_dist}(b). In this case, the energy shift coefficient $S_0$ from Eq.~(\ref{eq:E_shift}) can be estimated as:
\begin{equation}
	S_0 = \frac{r}{\Delta E} \left| \sum_{n=-\infty}^{+\infty} \frac{e^{i \frac{\Delta E d}{2\textcolor{black}{v_g}} n}}{\left( \frac{\epsilon - E_l}{\Delta E} \right) - n} \right|,
\end{equation}where $r \approx |\psi_n(\pm d/2)|$ denotes the approximate amplitude of the wavefunction at the connection sites, and $\textcolor{black}{v_g} = \frac{\partial E}{\partial k_x}$ is the group velocity along the longitudinal direction. For arbitrary real values of $\alpha = \frac{\Delta E d}{2 \textcolor{black}{v_g}}$ and $z = \frac{\epsilon - E_l}{\Delta E}$, the following identity can be established:
\begin{equation}
	\left| \sum_{n=-\infty}^{+\infty} \frac{e^{i \alpha n}}{z - n} \right| =
	\begin{cases}
		\frac{\pi}{\sin(\pi z)} & \text{for } \alpha \neq 0, \\\\
		\pi \cot(\pi z) & \text{for } \alpha = 0.
	\end{cases}
\end{equation}
This implies that any physical quantities derived from $S_0$, such as the effective oscillation frequency $\Omega_{\mathrm{eff}}$, become effectively independent of the separation distance $d$, except in the regime of small $d \sim 1$. In that limit, the condition $\alpha \to 0$ causes the summation to become sensitive to the discrete nature and finite count of contributing eigenstates, making the simplified approximation less reliable. This behavior is qualitatively supported by Fig.~\ref{fig:depend_dist}, which shows that both the oscillation fidelity and the effective frequency remain constant beyond a small threshold in $d$.\begin{figure}[t]
	\centering
	\includegraphics[width=0.95\linewidth]
    {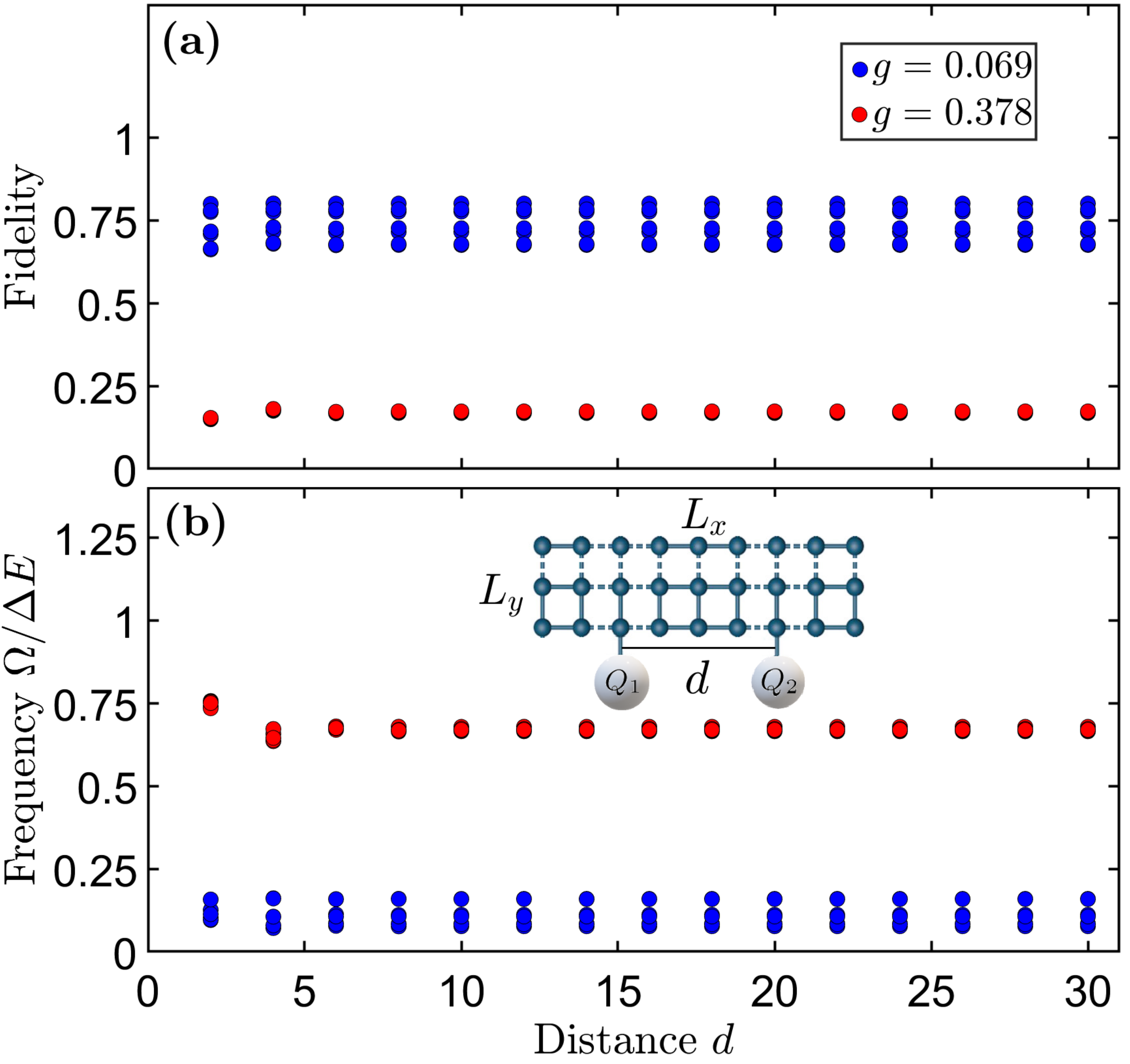}
	\caption{\label{fig:depend_dist}Induced interactions between two qubits symmetrically coupled to the longer edge of a rectangular Hofstadter lattice of dimensions $L_x = 105$ and $L_y = 21$. The connection sites are located at a distance of $\pm d/2$ from the central edge site, as illustrated in the (b) inset. Panels (a) and (b) show, respectively, the fidelity and effective oscillation frequency as functions of the distance $d$, calculated for two coupling strengths: $g = 0.069$ ($g^2 L/J^2 = 0.1$) and $g = 0.378$ ($g^2 L/J^2 = 3.0$). For each distance, values are obtained for 6 equally spaced values of $\epsilon$ within the interval $[E_l, E_{l+1}]$, with $E_l = -1.78$ and $E_{l+1} = -1.75$.}
\end{figure}

\textcolor{black}{\section{Experimental Setup}\label{sec_experiment}
Based on the results of this work, we can make an estimate of realistic experimental parameters, which include a lattice size of at least \( L = 7 \) and a hopping amplitude of \( J \approx 2\pi \times 18\,\text{MHz} \)~\cite{Owens2022}. The qubit-lattice coupling can range from a small percentage of $J$ (weak coupling) $g \approx 2\pi \times 0.2\,\text{MHz}$  up to several times \( J \) (strong coupling) $g \approx 2\pi \times 50\,\text{MHz}$. For the system size $L = 7$, that would result in effective frequencies \( \Omega_{\mathrm{eff}} \) ranging from approximately \( 2\pi \times 2\,\text{kHz} \) to \( 2\pi \times 5\,\text{MHz} \).  
A coupling strength of \( g \approx 0.38J \) corresponds to the regime \( g^2 L / J^2 \sim 1 \), which results in an effective frequency of \( \Omega_{\mathrm{eff}} \sim 2\pi \times 1.6\,\text{MHz} \).
The optimal placement of the qubit potential \( \epsilon \) is near the center of the gap between bulk states, at approximately \( \epsilon \approx -1.75 J\).}

\textcolor{black}{Practical implementation requires careful consideration of experimental imperfections. When multiple qubits are involved, both the coupling strength $g$ and the local potential $\epsilon$ may vary between sites due to calibration inaccuracies.
Appendix~\ref{ap1} analyzes how these variations influence key dynamical parameters, including the oscillation frequency $\Omega_{\mathrm{eff}}$ and the \textit{transfer probability} $p_\mathrm{tr}$. The latter is defined as twice the time-averaged excitation probability of the second qubit, normalized by the fidelity.
\begin{equation}
    p_{\mathrm{tr}}=\frac{2}{F}\lim_{T \to \infty} 
    \frac{1}{T} \int_{0}^{T}    
    |\Psi(Q_2, t)|^2 dt\,.\label{eq_p_tr}
\end{equation}
In other words, it quantifies the portion of probability that oscillates between the qubits, within the limit set by the time-averaged probability that remains in the two-qubit manifold.}

\textcolor{black}{A reduction of $p_{\mathrm{tr}}$ below unity is caused by a mismatch $\Delta\epsilon$ in the local potentials of the two qubits. Equation ~\eqref{eq:Kpr_g} provides an estimate of this effect and shows how it can be partially compensated by increasing the coupling strength $g$.
However, increasing $g$ comes at the cost of reduced fidelity, as shown in Eq.~\eqref{eq:fidel_anlt}, which leads to a uniform decrease in probability for both qubits.}

\textcolor{black}{For example, in the weak-to-moderate coupling regime \( g \approx 0.2J \) with a fidelity of approximately 75\% (assuming $L = 7$), introducing a small \( \Delta\epsilon \approx 0.08J \) reduces the transfer probability $p_\mathrm{tr}$ from 1 to 0.2. Boosting the coupling strength to \( g \approx 0.3J \) in an attempt to counteract this decay raises $p_{\mathrm{tr}}$ to about 0.56, though the fidelity subsequently drops to around 60\%.}

\textcolor{black}{
The effect of moderate disorder in the hopping amplitude \( J \) or the on-site potential is expected to be minimal, owing to the robustness of unidirectional, topologically protected modes against backscattering.  
Figure~\ref{fig:j_disorder}(a) shows that even with relatively strong disorder, at the level of 20\%, the energy gap remains open, and the edge mode appears largely unaffected within the relevant potential range \( -J \le \epsilon \le -2.5J \).
}

\textcolor{black}{
However, the impact of the disorder is not entirely negligible, as discussed in Appendix~\ref{ap1}. It may lead to a reduced transfer probability $p_\mathrm{tr}<1$, particularly if the parity symmetry described in Eq.~\eqref{eq:glob_symm} is not preserved.  
Equation~\eqref{eq:req_delta_eps} illustrates how these dressed energy levels can be brought back into resonance by introducing a small difference in the qubit potentials, denoted as \( \Delta \epsilon \). Eq.~\eqref{eq:Delta_eps_K_pr} provides an estimate of its required magnitude.
}

\textcolor{black}{Figures~\ref{fig:j_disorder}(b,c) demonstrate the effect of disorder on $p_{\mathrm{tr}}$, and how a small detuning of the second qubit potential \( \epsilon_2 \) can restore the desired behavior.  
In practice, due to the unpredictable nature of disorder, the exact value and sign of \( \Delta \epsilon \) will be need to be tuned experimentally.}\begin{figure}[t]
	\centering
	\includegraphics[width=0.95\linewidth]
    {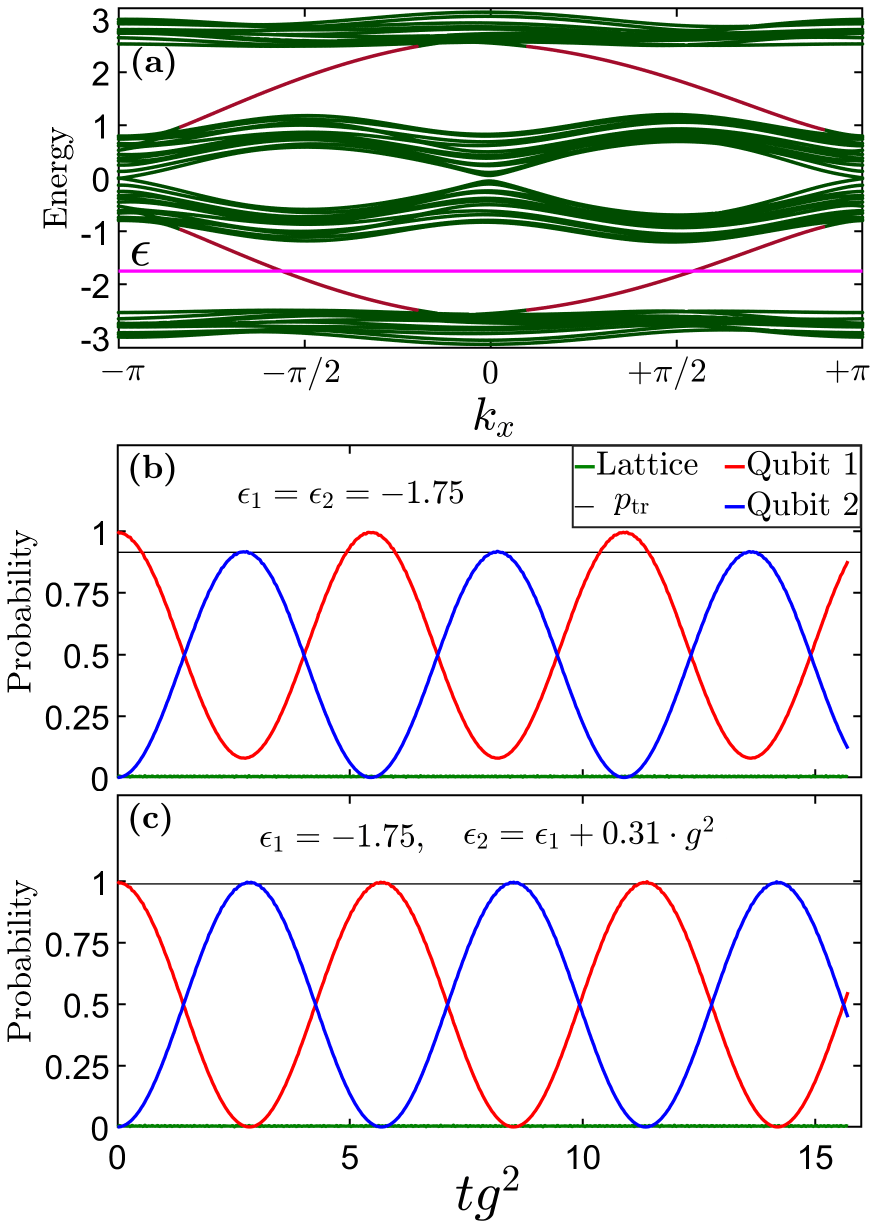}
   	\caption{ \label{fig:j_disorder}
    \textcolor{black}{
(a) Band structure of the Hofstadter lattice in a strip geometry ($L_y = 23$, $L_x \to \infty$), in the presence of strong columnar disorder in the hopping amplitude $J$ which takes random values in the interval $J \in [0.8, 1.2]$. Results shown are for a single disorder realization. Despite the disorder, the band structure clearly exhibits edge modes (red) crossing the large energy gaps between the three continuous bulk bands (green). The qubit energy $\epsilon \approx -1.75$ (magenta) remains within the edge mode spectrum.  (b, c) Time evolution of the qubit probabilities $P_{Q_1} = |\psi(Q_1,t)|^{2}$, $P_{Q_2} = |\psi(Q_2,t)|^{2}$, and the total lattice population $P_\mathrm{lat} = \sum_{j} |\psi(j,t)|^{2}$ for $L = 23$ and disordered hopping amplitudes $J$ taking random values in the interval $J \in [0.8, 1.2]$. Parameters: $g = 0.01$, first qubit potential $\epsilon_1 = -1.75$. Results shown are for a single disorder realization.
(b) The second qubit potential is set equal to the first, $\epsilon_2 = \epsilon_1$. A reduced transfer probability $p_\mathrm{tr} \approx 0.91 < 1$ is observed, caused by the disorder.
(c) The transfer probability is restored to $p_{\mathrm{tr}} \approx 1.0$ by introducing a small shift in the second qubit potential: $\epsilon_2 = \epsilon_1 + 0.31 \cdot g^2$.}}
\end{figure}

\section{Conclusions and outlook}\label{sec_concl}
The effect studied in this work exhibits several parallels with the Ruderman–Kittel–Kasuya–Yosida (RKKY) interaction \cite{10.1143/PTP.16.45,* PhysRev.106.893, *PhysRev.96.99}, where conduction electrons mediate indirect exchange. In our model, these electrons are analogous to electromagnetic resonators (sites). The crucial distinction, however, lies in the nature of the mediating states: here, the interaction is facilitated by topologically protected edge modes, rendering it virtually independent of distance. We have demonstrated that topological edge states in a Hofstadter lattice can mediate robust, long-range interactions between two localized qubits coupled to the system’s boundary. This mechanism could potentially be realized in other systems hosting topological edge states, such as spinful impurities coupled to the edge of a Hall state \cite{van_Dalum_2021, McGinley_2021}, chiral modes in anomalous Floquet insulators \cite{PhysRevB.99.195133, PhysRevX.6.021013}
or helical edge modes \cite{Rout2024, PhysRevB.87.165440}. The latter is particularly promising, having already found applications in technologies like dissipationless quantum spin 
transport \cite{doi:10.1126/science.1087128, doi:10.1126/science.1174736, Brüne2012}.

It is worth noting that similar results would be obtained from non-chiral one-dimensional photonic modes, although the interference of left- and right-movers would give an additional oscillation of the coupling amplitude with wave number $2 k_m$ for low-energy modes at $\pm k_m$. However, obtaining long-range coupling with such 1D models is challenging because it requires low enough disorder to reach the ballistic regime. This is formally challenging due to the instability of 1D systems to Anderson localization \cite{PhysRev.109.1492, Kramer1993}. Furthermore, it remains practically challenging in solid state devices; for example, ballistic nanowires are generally limited to submicron length scales \cite{Chuang2013, Kumar2024}. By contrast, quantum Hall states given quantized (Hall) transport over much larger distances.

Potential implementations include architectures where multiple transmons are coupled to a common lattice acting as a topological medium. Such setups would allow remote interactions mediated by protected edge states, enabling robust quantum information transmission. These interactions could serve as quantum links \cite{PRXQuantum.5.020363} in modular quantum processors. Furthermore, the inherent directionality of the coupling naturally simulates non-Hermitian dynamics \cite{Reisenbauer2024}, offering a platform for modeling dissipative quantum systems and synthetic gauge fields.

A potentially promising direction for further study is the non-perturbative regime, where the parameter combination enters the domain $g^2 L / J^2 \sim 1$\textcolor{black}{.} In this regime, the oscillation frequency becomes significantly less sensitive to the detuning of $\epsilon$ relative to the lattice eigenvalues, while the fidelity remains reasonably high ($\sim 0.5$), suggesting potential for practical applications.

Future research should investigate many-body scenarios involving multiple qubits coupled to the edge. These may include configurations where all qubits share a common potential, as well as architectures that isolate specific qubit pairs into protected communication channels by tuning them to distinct energy levels. A natural extension would be to generalize the current analysis beyond the single-excitation regime, exploring dynamics in the presence of multiple simultaneous excitations.

\section*{Acknowledgments}
We thank Andrei Vrajitoarea and Alicia Koll\'ar for valuable discussions. This work was performed with support from the National Science Foundation through award numbers DMR-1945529 and MPS-2228725 (M.K. and W.X.) and from the Welch Foundation through Award No. AT-2036-20200401. Part of this work was performed at the Aspen Center for Physics, which is supported by
NSF grant number PHY-1607611. M. Y. was supported at Georgetown by the Department of Energy, Office of Basic Energy Sciences, Division of Materials Sciences and Engineering under Contract No. DE-FG02-08ER46542 for writing the manuscript.

\onecolumngrid
\appendix

\section{Solution for quarter-flux Hofstadter lattice energy spectrum}\label{ap_lat}

Assuming that all blue sites in the Hofstadter lattice are located at positions where both the row $r$ and column $c$ are even, the most common form of the Schrödinger equation for bulk sites $(r,c)$ is given by:
\begin{subequations}
\begin{align}
-iJ\, \psi(r+1, c) + iJ\, \psi(r-1, c) + (-1)^r J\, \psi(r, c-1) - J\, \psi(r, c+1) &= \lambda\, \psi(r, c) && \text{for even } c, \label{eq:comm_schrod1} \\
-J\, \psi(r+1, c) - J\, \psi(r-1, c) + (-1)^r J\, \psi(r, c+1) - J\, \psi(r, c-1) &= \lambda\, \psi(r, c) && \text{for odd } c. \label{eq:comm_schrod2}
\end{align}
\end{subequations}

The bulk energy spectrum can be estimated by approximating the Hofstadter lattice as an infinite periodic structure composed of elementary 4-site square unit cells. Each unit cell consists of one blue site located at $(r_b, c_b)$ and three black sites located at $(r_b + 1, c_b)$, $(r_b, c_b + 1)$, and $(r_b + 1, c_b + 1)$. By applying Bloch's theorem and enforcing the periodic relationship 
\begin{equation}
\psi(r, c + 2) = \psi(r, c) e^{i k_x},\ \ \ \ \ \ \psi(r + 2, c) = \psi(r, c) e^{i k_y}, 
\end{equation}
on Eqs. \eqref{eq:comm_schrod1} and \eqref{eq:comm_schrod2}, 
we obtain the following secular equation and corresponding expression for the eigenvalue $\lambda$:
\begin{equation}
\Big(\frac{\lambda}{2J}\Big)^{4}-2\Big(\frac{\lambda}{2J}\Big)^{2}+\frac{1}{4}(\sin^{2}(k_{x})+\sin^{2}(k_{y}))=0 \implies
\Big(\frac{\lambda}{2J}\Big)^{2}=1\pm\sqrt{1-\frac{\sin^{2}(k_{x})+\sin^{2}(k_{y})}{4}}\,.
\end{equation}
Since $\sin^2(k_x)$ and $\sin^2(k_y)$ can vary between $0$ to $1$, we obtain the allowed ranges for the bulk energies:
\begin{equation}
0 \le \left|\frac{\lambda}{2J}\right| \le \sqrt{1 - \sqrt{\tfrac{1}{2}}}
\ \ \ \ \
\text{or}
\ \ \ \ \
		\sqrt{1 + \sqrt{\tfrac{1}{2}}} \le  \left|\frac{\lambda}{2J}\right| 
		\le \sqrt{2}\,.
\end{equation}

A different class of solutions, applicable to edge-mode states, emerges when considering the lattice in the strip geometry approximation, where \(L_y\) is finite and \(L_x \to \infty\). We consider a solution of the form:
\begin{align}
\psi(r,c) =
\begin{cases}
a_r \, e^{i\frac{k_{x}}{2} c} \cdot e^{\frac{\pi}{2}(r+1)} & \text{for even } c, \\
a_r \, e^{i\frac{k_{x}}{2} c} & \text{for odd } c.
\end{cases}
\end{align}
This trial function satisfies both Eqs.~\eqref{eq:comm_schrod1} and \eqref{eq:comm_schrod2} provided that the coefficients \(a_{r}\) satisfy the following recurrence relation:
\begin{equation} \label{eq:bethe}
-(a_{r+1} + a_{r-1}) = a_{r} \left( \frac{\lambda}{J} - 2\sin\left( \frac{k_x}{2} + \frac{\pi}{2}(r - r_0) \right) \right),
\end{equation}
which can, in principle, be solved using the Bethe Ansatz technique \cite{doi:10.1142/9789812795755_0004}. Here, $r_0$ reflects the ambiguity in defining the reference row $r = 0$. 
For large lattice sizes, it is convenient to assume that the vertical size of the system follows the pattern $L_y = 4n - 1$, where $n$ is a positive integer. This choice yields a particularly simple solution to Eq.~\eqref{eq:bethe}. 
In this case, the boundary conditions ($a_0 = a_{L_y + 1} = 0$) can be satisfied by setting every fourth row coefficient to zero, specifically, $a_{4k} = 0$ for $k = 0, 1, 2, \dots$. 
Equation \eqref{eq:bethe} allows for four distinct choices of $r_0 = 0, 1, 2, 3$.
Among these, the cases $r_0 = 0$ and $r_0 = 2$ correspond to the lattice configuration shown in Fig.~\ref{fig:qbt_hofst_lat}, where the edges are formed by black sites.
These two choices lead to the following characteristic equation for $\lambda$, along with the corresponding decay factor $d_r$:
\begin{equation}
    2\frac{\lambda}{J}-((\frac{\lambda}{J})^{2}-4\cos^{2}(\frac{k_x}{2}))(\frac{\lambda}{J} \pm 2\sin(\frac{k_x}{2}))=0 \label{eq:edge_lamb1}\,,
    \ \ \ \ \
    d_r =  \frac{a_{5}}{a_{1}}=-\frac{\frac{\lambda}{J} \mp 2\cos(\frac{k_x}{2})}{\frac{\lambda}{J} \pm 2\cos(\frac{k_x}{2})}\, ,
\end{equation}
where $d_r$ explicitly reveals the formation of unidirectional edge states.
The magnitude $|d_r|$ determines whether the state decays $(|d_r| < 1)$ or grows $(|d_r| > 1)$ in the $y$-direction, depending on the values of $k_x$ and $\lambda$. 
Equation~\eqref{eq:edge_lamb1} admits six distinct edge-mode branches. For any given $k_x$, three modes are localized at the lower edge and decay upward (in the positive $y$-direction), while the other three are localized at the upper edge and decay downward.

If the blue sites were located on the first row instead --- corresponding to the choices $r_0 = 1$ and $r_0 = 3$ --- the resulting relations for $\lambda$ and the decay factor $d_r$ would be different and given by:
\begin{equation}
    2\frac{\lambda}{J}-((\frac{\lambda}{J})^{2}-4\sin^{2}(\frac{k_x}{2}))(\frac{\lambda}{J} \pm 2\cos(\frac{k_x}{2}))=0 \label{eq:edge_lamb2}\,,
    \ \ \ \ \
    d_r =  \frac{a_{5}}{a_{1}}=-\frac{\frac{\lambda}{J} \mp 2\sin(\frac{k_x}{2})}{\frac{\lambda}{J} \pm 2\sin(\frac{k_x}{2})}\, .
\end{equation}

\section{Solution for asymmetrical coupling constants and potentials}\label{ap1}

In the case where the qubit potentials and coupling constants differ (denoted as $\epsilon_{1}$, $\epsilon_{2}$ and $g_{1}$, $g_{2}$, respectively), a solution can be constructed as a superposition of symmetric and antisymmetric eigenstates $\psi_{+}(j)$ and $\psi_{-}(j)$ for a given $\lambda$: $\psi(j) = a \psi_{+}(j) + b \psi_{-}(j)$. Substituting into Eqs.~\eqref{eq:Schrod_1_a} and~\eqref{eq:Schrod_1_b} for the two-qubit boundaries yields new analogous of~\eqref{eq:bc_4}, which now split into two separate equations:
\begin{subequations}
	\begin{align}
	\frac{g_{1}^{2}}{\lambda - \epsilon_{1}} &= \frac{\textcolor{black}{1}}{1 + \chi} \left( \frac{1}{f_{+}(\lambda)} + \frac{\chi}{f_{-}(\lambda)} \right), \label{eq:X_plus} \\
	\frac{g_{2}^{2}}{\lambda - \epsilon_{2}} &= \frac{\textcolor{black}{1}}{1 - \chi} \left( \frac{1}{f_{+}(\lambda)} - \frac{\chi}{f_{-}(\lambda)} \right), \label{eq:X_minus}
\end{align}\end{subequations}
where the dimensionless parameter $\chi$ accounts for normalization ambiguity $\chi = \frac{b}{a} \cdot \frac{\psi_{-}(1)}{\psi_{+}(1)}$. Solving Eqs.~\eqref{eq:X_plus} and \eqref{eq:X_minus} yields two solutions $\lambda_{1,2}$ and corresponding $\chi_{1,2}$, typically located near $\epsilon_{1}$ and $\epsilon_{2}$. These define the dominant eigenstates governing the coherent oscillation between the qubits.

Assuming the couplings $g_1$ and $g_2$ are weak enough that other states do not significantly contribute to the initial condition $\Psi(Q_1,t=0)=1$, we write the time evolution as:
\begin{equation}
	\Psi(j,t) = c_{1} \psi_{1}(j) e^{-i \lambda_{1} t} + c_{2} \psi_{2}(j) e^{-i \lambda_{2} t}.
\end{equation}
At $t=0$, only the first qubit is populated:
\begin{subequations}\begin{align}
	\Psi(Q_1, 0) &= \frac{-g_1}{\lambda_1 - \epsilon_1} c_1 a_1 \psi_{1+}(1)(1 + \chi_1) + \frac{-g_1}{\lambda_2 - \epsilon_1} c_2 a_2 \psi_{2+}(1)(1 + \chi_2) = 1, \\
	\Psi(Q_2, 0) &= \frac{-g_2}{\lambda_1 - \epsilon_2} c_1 a_1 \psi_{1+}(1)(1 - \chi_1)  + \frac{-g_2}{\lambda_2 - \epsilon_2} c_2 a_2 \psi_{2+}(1)(1 - \chi_2) = 0.
\end{align}\end{subequations}
Solving for $c_1$ and $c_2$ gives:
\begin{subequations}\begin{align}
	c_2 a_2 \psi_{2+}(1) &= -c_1 a_1 \psi_{1+}(1)
	\left( \frac{\lambda_2 - \epsilon_2}{1 - \chi_2} \right) \left( \frac{1 - \chi_1}{\lambda_1 - \epsilon_2} \right), \\
	c_1 a_1 \psi_{1+}(1) &= \frac{1}{g_1} \Bigg[
	\left( \frac{1 + \chi_2}{\lambda_2 - \epsilon_1} \right)
	\left( \frac{\lambda_2 - \epsilon_2}{1 - \chi_2} \right)
	\left( \frac{1 - \chi_1}{\lambda_1 - \epsilon_2} \right)  - \left( \frac{1 + \chi_1}{\lambda_1 - \epsilon_1} \right)
	\Bigg]^{-1}.
\end{align}\end{subequations}
At time $t_1 = \pi / |\lambda_2 - \lambda_1|$, the amplitude at Q$_2$ \textcolor{black}{reaches its maximum:}
\begin{align}
	|\Psi(Q_2, t_1)| = \left| \frac{2g_2}{g_1} \left[
	\left( \frac{1 + \chi_2}{\lambda_2 - \epsilon_1} \right)
	\left( \frac{\lambda_2 - \epsilon_2}{1 - \chi_2} \right) - \left( \frac{\lambda_1 - \epsilon_2}{1 - \chi_1} \right)
	\left( \frac{1 + \chi_1}{\lambda_1 - \epsilon_1} \right)
	\right]^{-1} \right|. \label{eq:Kpr}
\end{align}
\textcolor{black}{Within the perturbative limit covered in this section, the fidelity is assumed to be approximately unity, and therefore the average occupation probability of the second qubit is approximately one half of its maximum occupation probability. Then, using the definition of the transfer probability given by Eq.~\eqref{eq_p_tr}, we obtain $p_\mathrm{tr}$ as}
\begin{align}
	\textcolor{black}{p_{\mathrm{tr}}} & \textcolor{black}{\approx |\max \Psi(Q_2)|^{2}} = \frac{4g_2^2}{g_1^2} \left| 
	\left( \frac{1 + \chi_2}{\lambda_2 - \epsilon_1} \right)
	\left( \frac{\lambda_2 - \epsilon_2}{1 - \chi_2} \right) - \left( \frac{\lambda_1 - \epsilon_2}{1 - \chi_1} \right)
	\left( \frac{1 + \chi_1}{\lambda_1 - \epsilon_1} \right)
	\right|^{-2}.
\end{align}
Eliminating $\chi$ from Eqs.~\eqref{eq:X_plus} and \eqref{eq:X_minus} gives a general equation for $\lambda$:
\begin{align}
	&\frac{2g_1^2 g_2^2}{(\lambda - \epsilon_1)(\lambda - \epsilon_2)} + \frac{2}{f_{+}(\lambda) f_{-}(\lambda)} - \left( \frac{1}{f_{+}(\lambda)} + \frac{1}{f_{-}(\lambda)} \right)
	\left( \frac{g_1^2}{\lambda - \epsilon_1} + \frac{g_2^2}{\lambda - \epsilon_2} \right) = 0.
\end{align}
To simplify further, assume $f_{+}(\lambda)$ and $f_{-}(\lambda)$ vary slowly compared to steep lines \textcolor{black}{ $(\lambda - \epsilon_{1,2}) / g_{1,2}^2$}, allowing $f_{+}(\lambda) \approx \mathrm{const} = f_{+}$ and	$f_{-}(\lambda) \approx \mathrm{const} = f_{-}$. Then, Eq.~\eqref{eq:X_plus} and Eq.~\eqref{eq:X_minus} give:
\begin{align}
	\lambda = \frac{g_1^2(1 + \chi)}{\textcolor{black}{\left( \frac{1}{f_{+}} + \frac{\chi}{f_{-}} \right)}} + \epsilon_1 = \frac{g_2^2(1 - \chi)}{\textcolor{black}{\left( \frac{1}{f_{+}} - \frac{\chi}{f_{-}} \right)}} + \epsilon_2. \label{eq:Le1e2}
\end{align}
Solving Eq.~\eqref{eq:Le1e2} for $\chi$ gives:
\begin{align}
	\chi &= \frac{1}{2} \Bigg[
	-\textcolor{black}{(g_2^2 + g_1^2)}(f_{-} - f_{+}) \pm \sqrt{ 
		\left[ 2(\epsilon_2 - \epsilon_1) + \textcolor{black}{(g_2^2 - g_1^2)}(f_{-} + f_{+}) \right]^2 
		+ 4 \textcolor{black}{g_1^2 g_2^2} (f_{-} - f_{+})^2 }
	\Bigg] \nonumber \\
	&\quad \times \left[ \textcolor{black}{(g_2^2 - g_1^2)} f_{+} + (\epsilon_2 - \epsilon_1)\frac{f_{+}}{f_{-}} \right]^{-1}. \label{eq:X_sol_ex}
\end{align}
From these results, two key quantities can be extracted: the effective frequency,
\begin{align}
	\Omega_{\mathrm{eff}} = |\lambda_2 - \lambda_1| =\sqrt{ \left[ (\epsilon_2 - \epsilon_1) + \frac{1}{2} \textcolor{black}{\left( g_2^2 - g_1^2 \right)}(f_{-} + f_{+}) \right]^2 
		+ \textcolor{black}{g_1^2 g_2^2}(f_{-} - f_{+})^2 },
\end{align}
and the \textcolor{black}{transfer probability},
\begin{equation}
	\textcolor{black}{p_{\mathrm{tr}}} = \left| \frac{g_1 g_2 (f_{-} - f_{+})}{\textcolor{black}{\Omega_{\mathrm{eff}}}} \right|^2.
\end{equation}
The limiting cases can be achieved as 
\begin{itemize}
\item Symmetric case: $\epsilon_2 = \epsilon_1 = \epsilon$, $g_1 = g_2 = g$
\begin{align}
	\Omega_{\mathrm{eff}} = \textcolor{black}{g^2} |f_{-} - f_{+}|, \qquad
	\textcolor{black}{p_{\mathrm{tr}}} = 1.
\end{align}
\item Weak coupling limit: $g_{1,2}^2/J \ll |\epsilon_2 - \epsilon_1|$
\begin{align}
	\Omega_{\mathrm{eff}} \approx |\epsilon_2 - \epsilon_1|, \qquad
    \textcolor{black}{p_{\mathrm{tr}}} \approx \left| \frac{g_1 g_2 (f_{-} - f_{+})}{\textcolor{black}{|\epsilon_2 - \epsilon_1|}} \right|^2.
\end{align}
\item Symmetric coupling: $g_1 = g_2 = g$, $\epsilon_2 \neq \epsilon_1$
\begin{align}
\Omega_{\mathrm{eff}} = \sqrt{ (\epsilon_2 - \epsilon_1)^2 + \textcolor{black}{g^4}(f_{-} - f_{+})^2 },  \qquad
\textcolor{black}{p_{\mathrm{tr}}} = \left[ 1 + \frac{ \textcolor{black}{(\epsilon_2 - \epsilon_1)^2}}{g^4 (f_{-} - f_{+})^2} \right]^{-1},\qquad
g \ge \sqrt{ \frac{\textcolor{black}{|\epsilon_2 - \epsilon_1|}}{|f_{-} - f_{+}|} } \quad \text{for } \textcolor{black}{p_{\mathrm{tr}}} \ge \frac{1}{2} \label{eq:Kpr_g}.
\end{align}
\item Symmetric potentials: $\epsilon_1 = \epsilon_2 = \epsilon$, $g_1 \neq g_2$
\begin{align}
\Omega_{\mathrm{eff}} = \sqrt{
	\left[ \frac{1}{2} \textcolor{black}{\left( g_2^2 - g_1^2 \right)}(f_{-} + f_{+}) \right]^2 
	+ \textcolor{black}{g_1^2 g_2^2}(f_{-} - f_{+})^2 }, \qquad
\textcolor{black}{p_{\mathrm{tr}}} = \left[ 1 + \frac{1}{4} \left( \frac{g_2^2 - g_1^2}{g_1 g_2} \right)^2 
\left( \frac{f_{-} + f_{+}}{f_{-} - f_{+}} \right)^2 \right]^{-1}.
\end{align}
\end{itemize}

\textcolor{black}{
It should be noted that the correct zero-order states associated with the degenerate energy level $\epsilon$ may differ from those described in the idealized,
symmetric case given by Eq.~\eqref{eq:corr_split1}. This deviation can arise because the quantities $S_{1}(\epsilon)$ and $S_{2}(\epsilon)$, defined in Eqs.~\eqref{eq:abc-1}, are generally not equal.
Such an imbalance may occur, for example, if the connecting sites differ from the corner sites $1$ and $L^2$, or if there are local variations in the hopping coefficient $J$
that break the parity symmetry assumed in Eq.~\eqref{eq:glob_symm}. The general (non-normalized) form of the correct zero-order states in this case is:
\begin{equation}\label{eq:corr_states_gen}
|\phi_{q\pm}\rangle = e^{i\Delta \phi} |Q_{1}\rangle + \left( \gamma \pm \sqrt{\gamma^2 + 1} \right) |Q_{2}\rangle\,,\ \ \mathrm{where}\  \ \Delta \phi = \arg(S_{0}(\epsilon))\,,
\  \ \gamma = \frac{S_{1}(\epsilon) - S_{2}(\epsilon)}{|2 S_{0}(\epsilon)|}
\end{equation}
which leads to a value of the transfer probability less than $1$:
\begin{equation}\label{eq:K_pr_gamma}
p_{\mathrm{tr}} = \frac{1}{\gamma^2 + 1}\,.
\end{equation}
At the same time, the symmetric relationship of the zero-order states $\psi(Q_1) = \pm \psi(Q_2)$ can be effectively restored if the small coupling $g$ is accompanied by a corresponding small detuning of the qubit potential:
\begin{equation}
\Delta \epsilon = \epsilon_{2} - \epsilon_{1} = (S_{1}(\epsilon) - S_{2}(\epsilon)) g^2\,.\label{eq:req_delta_eps} 
\end{equation}
The required magnitude of $\Delta \epsilon$ can be directly estimated from the initially reduced value of $p_{\mathrm{tr}}$:
\begin{equation}\label{eq:Delta_eps_K_pr}
|\Delta\epsilon| = 2g^{2}|S_{0}(\epsilon)|\sqrt{\frac{1-p_{\mathrm{tr}}}{p_{\mathrm{tr}}}}\approx\Omega_{\mathrm{eff}}\sqrt{\frac{1 - p_{\mathrm{tr}}}{p_{\mathrm{tr}}}}
\end{equation}
}

\section{Construction of a symmetry operator on an odd-sized square Hofstadter lattice}\label{ap_sym}
This Appendix includes the development of the symmetry operator used in the derivation in Section \ref{sec_perturb}. As discussed in the main text, a rotation operator $R_\pi$ can be defined for a square Hofstadter lattice with an odd side length $L$:
\begin{align}
R_\pi = \sum_j | j \rangle \langle L^2 + 1 - j |\, .
\end{align}
It can also be applied to the lattice that is coupled to the qubits, as shown in Fig.~\ref{fig:qbt_hofst_lat}. This operator exchanges qubit $Q_1$ with qubit $Q_2$, maps the first lattice site $j=1$ to the last site $j^\prime =L^2$, and generally swaps each site $j$ with its mirror counterpart $j^\prime = L^2 + 1 -j$. The operator is unitary, satisfying $R_\pi R_\pi^\dagger = 1$.

Additionally, we introduce an operator
\begin{align}
S_b = \sum_j b(j) \   | j \rangle \langle j |\,,
\end{align}
which flips the signs of the blue sites. This operator is also unitary, with $S_b S_b^\dagger = 1$. The function $b(j)$ is defined as
\begin{align}
b(j) = \begin{cases}
-1& \text{for blue sites}, \\
+1& \text{for qubits and black sites}.
\end{cases}
\end{align}As discussed in the main text, Sec.~\ref{sec_perturb}, the states obtained by flipping the signs on the blue sublattice, $S_b |\psi_\lambda \rangle$, are eigenstates of the rotated Hamiltonian $R_\pi^\dagger H R_\pi$:
\begin{equation}
R_\pi^\dagger H R_\pi S_b |\psi_\lambda \rangle = \lambda S_b |\psi_\lambda \rangle \implies H R_\pi S_b |\psi_\lambda \rangle = \lambda R_\pi S_b |\psi_\lambda \rangle\,. \label{eq_H_U}
\end{equation}
On the other hand:
\begin{equation}
H |\psi_\lambda \rangle = \lambda |\psi_\lambda \rangle \implies R_\pi S_b H |\psi_\lambda \rangle = \lambda R_\pi S_b |\psi_\lambda \rangle\,. \label{eq_U_H}
\end{equation}
Since the relations \eqref{eq_H_U} and \eqref{eq_U_H} hold for any arbitrary eigenstate $|\psi_\lambda\rangle$, the following holds for any $|\psi\rangle$---superposition of eigenstates: $H (R_\pi S_b)|\psi\rangle = (R_\pi S_b) H|\psi\rangle$. This implies that the unitary symmetry operator $U_s = R_\pi S_b$ commutes with the Hamiltonian, i.e., $[U_s,H] = 0$ and $U_s|\psi_\lambda\rangle$ is also an eigenstate of $H$ with the same eigenvalue $\lambda$. Since $U_s^2 =1$, the eigenstates can be classified by their parity under $U_s$, with $|\psi_\lambda\rangle = \pm U_s |\psi_\lambda\rangle$. The explicit form of $U_s$ is:
\begin{equation}
U_s = \sum_j b(j)\  | j \rangle \langle L^2 + 1 - j|\,.
\end{equation}
}
\twocolumngrid
\bibliography{bib.bib}

\begin{thebibliography}{53}%
\makeatletter
\providecommand \@ifxundefined [1]{%
 \@ifx{#1\undefined}
}%
\providecommand \@ifnum [1]{%
 \ifnum #1\expandafter \@firstoftwo
 \else \expandafter \@secondoftwo
 \fi
}%
\providecommand \@ifx [1]{%
 \ifx #1\expandafter \@firstoftwo
 \else \expandafter \@secondoftwo
 \fi
}%
\providecommand \natexlab [1]{#1}%
\providecommand \enquote  [1]{``#1''}%
\providecommand \bibnamefont  [1]{#1}%
\providecommand \bibfnamefont [1]{#1}%
\providecommand \citenamefont [1]{#1}%
\providecommand \href@noop [0]{\@secondoftwo}%
\providecommand \href [0]{\begingroup \@sanitize@url \@href}%
\providecommand \@href[1]{\@@startlink{#1}\@@href}%
\providecommand \@@href[1]{\endgroup#1\@@endlink}%
\providecommand \@sanitize@url [0]{\catcode `\\12\catcode `\$12\catcode `\&12\catcode `\#12\catcode `\^12\catcode `\_12\catcode `\%12\relax}%
\providecommand \@@startlink[1]{}%
\providecommand \@@endlink[0]{}%
\providecommand \url  [0]{\begingroup\@sanitize@url \@url }%
\providecommand \@url [1]{\endgroup\@href {#1}{\urlprefix }}%
\providecommand \urlprefix  [0]{URL }%
\providecommand \Eprint [0]{\href }%
\providecommand \doibase [0]{https://doi.org/}%
\providecommand \selectlanguage [0]{\@gobble}%
\providecommand \bibinfo  [0]{\@secondoftwo}%
\providecommand \bibfield  [0]{\@secondoftwo}%
\providecommand \translation [1]{[#1]}%
\providecommand \BibitemOpen [0]{}%
\providecommand \bibitemStop [0]{}%
\providecommand \bibitemNoStop [0]{.\EOS\space}%
\providecommand \EOS [0]{\spacefactor3000\relax}%
\providecommand \BibitemShut  [1]{\csname bibitem#1\endcsname}%
\let\auto@bib@innerbib\@empty
\bibitem [{\citenamefont {{Kiczynski, M. and Gorman, S. K. and Geng, H. and Donnelly, M. B. and Chung, Y. and He, Y. and Keizer, J. G. and Simmons, M. Y.}}(2022)}]{Kiczynski2022}%
  \BibitemOpen
  \bibfield  {author} {\bibinfo {author} {\bibnamefont {{Kiczynski, M. and Gorman, S. K. and Geng, H. and Donnelly, M. B. and Chung, Y. and He, Y. and Keizer, J. G. and Simmons, M. Y.}}},\ }\bibfield  {title} {\bibinfo {title} {{Engineering topological states in atom-based semiconductor quantum dots}},\ }\href {https://doi.org/10.1038/s41586-022-04706-0} {\bibfield  {journal} {\bibinfo  {journal} {{Nature}}\ }\textbf {\bibinfo {volume} {606}},\ \bibinfo {pages} {694} (\bibinfo {year} {2022})}\BibitemShut {NoStop}%
\bibitem [{\citenamefont {{Zhang, Dan-Wei and Zhu, Yan-Qing and Zhao, Y. X. and Yan, Hui and Zhu, Shi-Liang}}(2018)}]{Zhang02102018}%
  \BibitemOpen
  \bibfield  {author} {\bibinfo {author} {\bibnamefont {{Zhang, Dan-Wei and Zhu, Yan-Qing and Zhao, Y. X. and Yan, Hui and Zhu, Shi-Liang}}},\ }\bibfield  {title} {\bibinfo {title} {{Topological quantum matter with cold atoms}},\ }\href {https://doi.org/10.1080/00018732.2019.1594094} {\bibfield  {journal} {\bibinfo  {journal} {{Advances in Physics}}\ }\textbf {\bibinfo {volume} {67}},\ \bibinfo {pages} {253} (\bibinfo {year} {2018})}\BibitemShut {NoStop}%
\bibitem [{\citenamefont {{Rachel, Stephan}}(2018)}]{Rachel_2018}%
  \BibitemOpen
  \bibfield  {author} {\bibinfo {author} {\bibnamefont {{Rachel, Stephan}}},\ }\bibfield  {title} {\bibinfo {title} {{Interacting topological insulators: a review}},\ }\href {https://doi.org/10.1088/1361-6633/aad6a6} {\bibfield  {journal} {\bibinfo  {journal} {{Reports on Progress in Physics}}\ }\textbf {\bibinfo {volume} {81}},\ \bibinfo {pages} {116501} (\bibinfo {year} {2018})}\BibitemShut {NoStop}%
\bibitem [{\citenamefont {{Ozawa, Tomoki and Price, Hannah M. and Amo, Alberto and Goldman, Nathan and Hafezi, Mohammad and Lu, Ling and Rechtsman, Mikael C. and Schuster, David and Simon, Jonathan and Zilberberg, Oded and Carusotto, Iacopo}}(2019)}]{RevModPhys.91.015006}%
  \BibitemOpen
  \bibfield  {author} {\bibinfo {author} {\bibnamefont {{Ozawa, Tomoki and Price, Hannah M. and Amo, Alberto and Goldman, Nathan and Hafezi, Mohammad and Lu, Ling and Rechtsman, Mikael C. and Schuster, David and Simon, Jonathan and Zilberberg, Oded and Carusotto, Iacopo}}},\ }\bibfield  {title} {\bibinfo {title} {{Topological photonics}},\ }\href {https://doi.org/10.1103/RevModPhys.91.015006} {\bibfield  {journal} {\bibinfo  {journal} {{Rev. Mod. Phys.}}\ }\textbf {\bibinfo {volume} {91}},\ \bibinfo {pages} {015006} (\bibinfo {year} {2019})}\BibitemShut {NoStop}%
\bibitem [{\citenamefont {{Nemirovsky, Liat and Cohen, Moshe-Ishay and Lumer, Yaakov and Lustig, Eran and Segev, Mordechai}}(2021)}]{PhysRevLett.127.093901}%
  \BibitemOpen
  \bibfield  {author} {\bibinfo {author} {\bibnamefont {{Nemirovsky, Liat and Cohen, Moshe-Ishay and Lumer, Yaakov and Lustig, Eran and Segev, Mordechai}}},\ }\bibfield  {title} {\bibinfo {title} {{Synthetic-Space Photonic Topological Insulators Utilizing Dynamically Invariant Structure}},\ }\href {https://doi.org/10.1103/PhysRevLett.127.093901} {\bibfield  {journal} {\bibinfo  {journal} {{Phys. Rev. Lett.}}\ }\textbf {\bibinfo {volume} {127}},\ \bibinfo {pages} {093901} (\bibinfo {year} {2021})}\BibitemShut {NoStop}%
\bibitem [{\citenamefont {{Bao, Changhua and Tang, Peizhe and Sun, Dong and Zhou, Shuyun}}(2022)}]{Bao2022}%
  \BibitemOpen
  \bibfield  {author} {\bibinfo {author} {\bibnamefont {{Bao, Changhua and Tang, Peizhe and Sun, Dong and Zhou, Shuyun}}},\ }\bibfield  {title} {\bibinfo {title} {{Light-induced emergent phenomena in 2D materials and topological materials}},\ }\href {https://doi.org/10.1038/s42254-021-00388-1} {\bibfield  {journal} {\bibinfo  {journal} {{Nature Reviews Physics}}\ }\textbf {\bibinfo {volume} {4}},\ \bibinfo {pages} {33} (\bibinfo {year} {2022})}\BibitemShut {NoStop}%
\bibitem [{\citenamefont {{Tan, Wei and Sun, Yong and Chen, Hong and Shen, Shun-Qing}}(2014)}]{Tan2014}%
  \BibitemOpen
  \bibfield  {author} {\bibinfo {author} {\bibnamefont {{Tan, Wei and Sun, Yong and Chen, Hong and Shen, Shun-Qing}}},\ }\bibfield  {title} {\bibinfo {title} {{Photonic simulation of topological excitations in metamaterials}},\ }\href {https://doi.org/10.1038/srep03842} {\bibfield  {journal} {\bibinfo  {journal} {{Scientific Reports}}\ }\textbf {\bibinfo {volume} {4}},\ \bibinfo {pages} {3842} (\bibinfo {year} {2014})}\BibitemShut {NoStop}%
\bibitem [{\citenamefont {{Rechtsman, Mikael C. and Zeuner, Julia M. and Plotnik, Yonatan and Lumer, Yaakov and Podolsky, Daniel and Dreisow, Felix and Nolte, Stefan and Segev, Mordechai and Szameit, Alexander}}(2013)}]{Rechtsman_2013}%
  \BibitemOpen
  \bibfield  {author} {\bibinfo {author} {\bibnamefont {{Rechtsman, Mikael C. and Zeuner, Julia M. and Plotnik, Yonatan and Lumer, Yaakov and Podolsky, Daniel and Dreisow, Felix and Nolte, Stefan and Segev, Mordechai and Szameit, Alexander}}},\ }\bibfield  {title} {\bibinfo {title} {{Photonic Floquet topological insulators}},\ }\href {https://doi.org/10.1038/nature12066} {\bibfield  {journal} {\bibinfo  {journal} {{Nature}}\ }\textbf {\bibinfo {volume} {496}},\ \bibinfo {pages} {196–200} (\bibinfo {year} {2013})}\BibitemShut {NoStop}%
\bibitem [{\citenamefont {{Carusotto, Iacopo and Ciuti, Cristiano}}(2013)}]{RevModPhys.85.299}%
  \BibitemOpen
  \bibfield  {author} {\bibinfo {author} {\bibnamefont {{Carusotto, Iacopo and Ciuti, Cristiano}}},\ }\bibfield  {title} {\bibinfo {title} {{Quantum fluids of light}},\ }\href {https://doi.org/10.1103/RevModPhys.85.299} {\bibfield  {journal} {\bibinfo  {journal} {{Rev. Mod. Phys.}}\ }\textbf {\bibinfo {volume} {85}},\ \bibinfo {pages} {299} (\bibinfo {year} {2013})}\BibitemShut {NoStop}%
\bibitem [{\citenamefont {{Krishnamoorthy, Harish N. S. and Dubrovkin, Alexander M. and Adamo, Giorgio and Soci, Cesare}}(2023)}]{Krishnamoorthy2023}%
  \BibitemOpen
  \bibfield  {author} {\bibinfo {author} {\bibnamefont {{Krishnamoorthy, Harish N. S. and Dubrovkin, Alexander M. and Adamo, Giorgio and Soci, Cesare}}},\ }\bibfield  {title} {\bibinfo {title} {{Topological Insulator Metamaterials}},\ }\href {https://doi.org/10.1021/acs.chemrev.2c00594} {\bibfield  {journal} {\bibinfo  {journal} {{Chemical Reviews}}\ }\textbf {\bibinfo {volume} {123}},\ \bibinfo {pages} {4416} (\bibinfo {year} {2023})}\BibitemShut {NoStop}%
\bibitem [{\citenamefont {{MacDonald, A. H.}}(1983)}]{PhysRevB.28.6713}%
  \BibitemOpen
  \bibfield  {author} {\bibinfo {author} {\bibnamefont {{MacDonald, A. H.}}},\ }\bibfield  {title} {\bibinfo {title} {{Landau-level subband structure of electrons on a square lattice}},\ }\href {https://doi.org/10.1103/PhysRevB.28.6713} {\bibfield  {journal} {\bibinfo  {journal} {{Phys. Rev. B}}\ }\textbf {\bibinfo {volume} {28}},\ \bibinfo {pages} {6713} (\bibinfo {year} {1983})}\BibitemShut {NoStop}%
\bibitem [{\citenamefont {{Padavi\ifmmode \acute{c}\else \'{c}\fi{}, Karmela and Hegde, Suraj S. and DeGottardi, Wade and Vishveshwara, Smitha}}(2018)}]{PhysRevB.98.024205}%
  \BibitemOpen
  \bibfield  {author} {\bibinfo {author} {\bibnamefont {{Padavi\ifmmode \acute{c}\else \'{c}\fi{}, Karmela and Hegde, Suraj S. and DeGottardi, Wade and Vishveshwara, Smitha}}},\ }\bibfield  {title} {\bibinfo {title} {{Topological phases, edge modes, and the Hofstadter butterfly in coupled Su-Schrieffer-Heeger systems}},\ }\href {https://doi.org/10.1103/PhysRevB.98.024205} {\bibfield  {journal} {\bibinfo  {journal} {{Phys. Rev. B}}\ }\textbf {\bibinfo {volume} {98}},\ \bibinfo {pages} {024205} (\bibinfo {year} {2018})}\BibitemShut {NoStop}%
\bibitem [{\citenamefont {{Scaffidi, Thomas and Simon, Steven H.}}(2014)}]{PhysRevB.90.115132}%
  \BibitemOpen
  \bibfield  {author} {\bibinfo {author} {\bibnamefont {{Scaffidi, Thomas and Simon, Steven H.}}},\ }\bibfield  {title} {\bibinfo {title} {{Exact solutions of fractional Chern insulators: Interacting particles in the Hofstadter model at finite size}},\ }\href {https://doi.org/10.1103/PhysRevB.90.115132} {\bibfield  {journal} {\bibinfo  {journal} {{Phys. Rev. B}}\ }\textbf {\bibinfo {volume} {90}},\ \bibinfo {pages} {115132} (\bibinfo {year} {2014})}\BibitemShut {NoStop}%
\bibitem [{\citenamefont {{Wang, Lei and Hung, Hsiang-Hsuan and Troyer, Matthias}}(2014)}]{PhysRevB.90.205111}%
  \BibitemOpen
  \bibfield  {author} {\bibinfo {author} {\bibnamefont {{Wang, Lei and Hung, Hsiang-Hsuan and Troyer, Matthias}}},\ }\bibfield  {title} {\bibinfo {title} {{Topological phase transition in the Hofstadter-Hubbard model}},\ }\href {https://doi.org/10.1103/PhysRevB.90.205111} {\bibfield  {journal} {\bibinfo  {journal} {{Phys. Rev. B}}\ }\textbf {\bibinfo {volume} {90}},\ \bibinfo {pages} {205111} (\bibinfo {year} {2014})}\BibitemShut {NoStop}%
\bibitem [{\citenamefont {{Herzog-Arbeitman, Jonah and Song, Zhi-Da and Regnault, Nicolas and Bernevig, B. Andrei}}(2020)}]{PhysRevLett.125.236804}%
  \BibitemOpen
  \bibfield  {author} {\bibinfo {author} {\bibnamefont {{Herzog-Arbeitman, Jonah and Song, Zhi-Da and Regnault, Nicolas and Bernevig, B. Andrei}}},\ }\bibfield  {title} {\bibinfo {title} {{Hofstadter Topology: Noncrystalline Topological Materials at High Flux}},\ }\href {https://doi.org/10.1103/PhysRevLett.125.236804} {\bibfield  {journal} {\bibinfo  {journal} {{Phys. Rev. Lett.}}\ }\textbf {\bibinfo {volume} {125}},\ \bibinfo {pages} {236804} (\bibinfo {year} {2020})}\BibitemShut {NoStop}%
\bibitem [{\citenamefont {{Hofstadter, Douglas R.}}(1976)}]{PhysRevB.14.2239}%
  \BibitemOpen
  \bibfield  {author} {\bibinfo {author} {\bibnamefont {{Hofstadter, Douglas R.}}},\ }\bibfield  {title} {\bibinfo {title} {{Energy levels and wave functions of Bloch electrons in rational and irrational magnetic fields}},\ }\href {https://doi.org/10.1103/PhysRevB.14.2239} {\bibfield  {journal} {\bibinfo  {journal} {{Phys. Rev. B}}\ }\textbf {\bibinfo {volume} {14}},\ \bibinfo {pages} {2239} (\bibinfo {year} {1976})}\BibitemShut {NoStop}%
\bibitem [{\citenamefont {{Owens, John Clai and Panetta, Margaret G. and Saxberg, Brendan and Roberts, Gabrielle and Chakram, Srivatsan and Ma, Ruichao and Vrajitoarea, Andrei and Simon, Jonathan and Schuster, David I.}}(2022)}]{Owens2022}%
  \BibitemOpen
  \bibfield  {author} {\bibinfo {author} {\bibnamefont {{Owens, John Clai and Panetta, Margaret G. and Saxberg, Brendan and Roberts, Gabrielle and Chakram, Srivatsan and Ma, Ruichao and Vrajitoarea, Andrei and Simon, Jonathan and Schuster, David I.}}},\ }\bibfield  {title} {\bibinfo {title} {{Chiral cavity quantum electrodynamics}},\ }\href {https://doi.org/10.1038/s41567-022-01671-3} {\bibfield  {journal} {\bibinfo  {journal} {{Nature Physics}}\ }\textbf {\bibinfo {volume} {18}},\ \bibinfo {pages} {1048} (\bibinfo {year} {2022})}\BibitemShut {NoStop}%
\bibitem [{\citenamefont {{Owens, Clai and LaChapelle, Aman and Saxberg, Brendan and Anderson, Brandon M. and Ma, Ruichao and Simon, Jonathan and Schuster, David I.}}(2018)}]{PhysRevA.97.013818}%
  \BibitemOpen
  \bibfield  {author} {\bibinfo {author} {\bibnamefont {{Owens, Clai and LaChapelle, Aman and Saxberg, Brendan and Anderson, Brandon M. and Ma, Ruichao and Simon, Jonathan and Schuster, David I.}}},\ }\bibfield  {title} {\bibinfo {title} {{Quarter-flux Hofstadter lattice in a qubit-compatible microwave cavity array}},\ }\href {https://doi.org/10.1103/PhysRevA.97.013818} {\bibfield  {journal} {\bibinfo  {journal} {{Phys. Rev. A}}\ }\textbf {\bibinfo {volume} {97}},\ \bibinfo {pages} {013818} (\bibinfo {year} {2018})}\BibitemShut {NoStop}%
\bibitem [{\citenamefont {{Het\'enyi, Bence and Mook, Alexander and Klinovaja, Jelena and Loss, Daniel}}(2022)}]{PhysRevB.106.235409}%
  \BibitemOpen
  \bibfield  {author} {\bibinfo {author} {\bibnamefont {{Het\'enyi, Bence and Mook, Alexander and Klinovaja, Jelena and Loss, Daniel}}},\ }\bibfield  {title} {\bibinfo {title} {{Long-distance coupling of spin qubits via topological magnons}},\ }\href {https://doi.org/10.1103/PhysRevB.106.235409} {\bibfield  {journal} {\bibinfo  {journal} {{Phys. Rev. B}}\ }\textbf {\bibinfo {volume} {106}},\ \bibinfo {pages} {235409} (\bibinfo {year} {2022})}\BibitemShut {NoStop}%
\bibitem [{\citenamefont {{Gu, Feng-Lei and Liu, Jia and Mei, Feng and Jia, Suotang and Zhang, Dan-Wei and Xue, Zheng-Yuan}}(2019)}]{Gu_2019}%
  \BibitemOpen
  \bibfield  {author} {\bibinfo {author} {\bibnamefont {{Gu, Feng-Lei and Liu, Jia and Mei, Feng and Jia, Suotang and Zhang, Dan-Wei and Xue, Zheng-Yuan}}},\ }\bibfield  {title} {\bibinfo {title} {{Synthetic spin–orbit coupling and topological polaritons in Janeys–Cummings lattices}},\ }\href {https://doi.org/10.1038/s41534-019-0148-9} {\bibfield  {journal} {\bibinfo  {journal} {{npj Quantum Information}}\ }\textbf {\bibinfo {volume} {5}},\ \bibinfo {pages} {36} (\bibinfo {year} {2019})}\BibitemShut {NoStop}%
\bibitem [{\citenamefont {{Mei, Feng and Chen, Gang and Tian, Lin and Zhu, Shi-Liang and Jia, Suotang}}(2018)}]{PhysRevA.98.012331}%
  \BibitemOpen
  \bibfield  {author} {\bibinfo {author} {\bibnamefont {{Mei, Feng and Chen, Gang and Tian, Lin and Zhu, Shi-Liang and Jia, Suotang}}},\ }\bibfield  {title} {\bibinfo {title} {{Robust quantum state transfer via topological edge states in superconducting qubit chains}},\ }\href {https://doi.org/10.1103/PhysRevA.98.012331} {\bibfield  {journal} {\bibinfo  {journal} {{Phys. Rev. A}}\ }\textbf {\bibinfo {volume} {98}},\ \bibinfo {pages} {012331} (\bibinfo {year} {2018})}\BibitemShut {NoStop}%
\bibitem [{\citenamefont {{Pakkiam, Prasanna and Kumar, N. Pradeep and Pletyukhov, Mikhail and Fedorov, Arkady}}(2023)}]{Pakkiam2023}%
  \BibitemOpen
  \bibfield  {author} {\bibinfo {author} {\bibnamefont {{Pakkiam, Prasanna and Kumar, N. Pradeep and Pletyukhov, Mikhail and Fedorov, Arkady}}},\ }\bibfield  {title} {\bibinfo {title} {{Qubit-controlled directional edge states in waveguide QED}},\ }\href {https://doi.org/10.1038/s41534-023-00722-8} {\bibfield  {journal} {\bibinfo  {journal} {{npj Quantum Information}}\ }\textbf {\bibinfo {volume} {9}},\ \bibinfo {pages} {53} (\bibinfo {year} {2023})}\BibitemShut {NoStop}%
\bibitem [{\citenamefont {{Mi, X. and Sonner, M. and Niu, M. Y. and Lee, K. W. and Foxen, B. and Acharya, R. and Aleiner, I. and Andersen, T. I. and Arute, F. and Arya, K. and others}}(2022)}]{doi:10.1126/science.abq5769}%
  \BibitemOpen
  \bibfield  {author} {\bibinfo {author} {\bibnamefont {{Mi, X. and Sonner, M. and Niu, M. Y. and Lee, K. W. and Foxen, B. and Acharya, R. and Aleiner, I. and Andersen, T. I. and Arute, F. and Arya, K. and others}}},\ }\bibfield  {title} {\bibinfo {title} {{Noise-resilient edge modes on a chain of superconducting qubits}},\ }\href {https://doi.org/10.1126/science.abq5769} {\bibfield  {journal} {\bibinfo  {journal} {{Science}}\ }\textbf {\bibinfo {volume} {378}},\ \bibinfo {pages} {785} (\bibinfo {year} {2022})}\BibitemShut {NoStop}%
\bibitem [{\citenamefont {{Groh, Thorsten and Brakhane, Stefan and Alt, Wolfgang and Meschede, Dieter and Asb\'oth, Janos K. and Alberti, Andrea}}(2016)}]{PhysRevA.94.013620}%
  \BibitemOpen
  \bibfield  {author} {\bibinfo {author} {\bibnamefont {{Groh, Thorsten and Brakhane, Stefan and Alt, Wolfgang and Meschede, Dieter and Asb\'oth, Janos K. and Alberti, Andrea}}},\ }\bibfield  {title} {\bibinfo {title} {{Robustness of topologically protected edge states in quantum walk experiments with neutral atoms}},\ }\href {https://doi.org/10.1103/PhysRevA.94.013620} {\bibfield  {journal} {\bibinfo  {journal} {{Phys. Rev. A}}\ }\textbf {\bibinfo {volume} {94}},\ \bibinfo {pages} {013620} (\bibinfo {year} {2016})}\BibitemShut {NoStop}%
\bibitem [{\citenamefont {{Kasuya, Tadao}}(1956)}]{10.1143/PTP.16.45}%
  \BibitemOpen
  \bibfield  {author} {\bibinfo {author} {\bibnamefont {{Kasuya, Tadao}}},\ }\bibfield  {title} {\bibinfo {title} {{A Theory of Metallic Ferro- and Antiferromagnetism on Zener's Model}},\ }\href {https://doi.org/10.1143/PTP.16.45} {\bibfield  {journal} {\bibinfo  {journal} {{Progress of Theoretical Physics}}\ }\textbf {\bibinfo {volume} {16}},\ \bibinfo {pages} {45} (\bibinfo {year} {1956})}\BibitemShut {NoStop}%
\bibitem [{\citenamefont {{Yosida, Kei}}(1957)}]{PhysRev.106.893}%
  \BibitemOpen
  \bibfield  {author} {\bibinfo {author} {\bibnamefont {{Yosida, Kei}}},\ }\bibfield  {title} {\bibinfo {title} {{Magnetic Properties of {C}u-{M}n Alloys}},\ }\href {https://doi.org/10.1103/PhysRev.106.893} {\bibfield  {journal} {\bibinfo  {journal} {{Phys. Rev.}}\ }\textbf {\bibinfo {volume} {106}},\ \bibinfo {pages} {893} (\bibinfo {year} {1957})}\BibitemShut {NoStop}%
\bibitem [{\citenamefont {{Ruderman, M. A. and Kittel, C.}}(1954)}]{PhysRev.96.99}%
  \BibitemOpen
  \bibfield  {author} {\bibinfo {author} {\bibnamefont {{Ruderman, M. A. and Kittel, C.}}},\ }\bibfield  {title} {\bibinfo {title} {{Indirect Exchange Coupling of Nuclear Magnetic Moments by Conduction Electrons}},\ }\href {https://doi.org/10.1103/PhysRev.96.99} {\bibfield  {journal} {\bibinfo  {journal} {{Phys. Rev.}}\ }\textbf {\bibinfo {volume} {96}},\ \bibinfo {pages} {99} (\bibinfo {year} {1954})}\BibitemShut {NoStop}%
\bibitem [{\citenamefont {{Li, Jia-Qi and Gao, Zhao-Min and Liu, Wen-Xiao and Wang, Xin}}(2023)}]{PhysRevA.108.043708}%
  \BibitemOpen
  \bibfield  {author} {\bibinfo {author} {\bibnamefont {{Li, Jia-Qi and Gao, Zhao-Min and Liu, Wen-Xiao and Wang, Xin}}},\ }\bibfield  {title} {\bibinfo {title} {{Light-matter interactions in a Hofstadter lattice with next-nearest-neighbor couplings}},\ }\href {https://doi.org/10.1103/PhysRevA.108.043708} {\bibfield  {journal} {\bibinfo  {journal} {{Phys. Rev. A}}\ }\textbf {\bibinfo {volume} {108}},\ \bibinfo {pages} {043708} (\bibinfo {year} {2023})}\BibitemShut {NoStop}%
\bibitem [{\citenamefont {{Berti, Anna and Carusotto, Iacopo}}(2022)}]{PhysRevA.105.023329}%
  \BibitemOpen
  \bibfield  {author} {\bibinfo {author} {\bibnamefont {{Berti, Anna and Carusotto, Iacopo}}},\ }\bibfield  {title} {\bibinfo {title} {{Topological two-particle dynamics in a periodically driven lattice model with on-site interactions}},\ }\href {https://doi.org/10.1103/PhysRevA.105.023329} {\bibfield  {journal} {\bibinfo  {journal} {{Phys. Rev. A}}\ }\textbf {\bibinfo {volume} {105}},\ \bibinfo {pages} {023329} (\bibinfo {year} {2022})}\BibitemShut {NoStop}%
\bibitem [{\citenamefont {{Irsigler, Bernhard and Zheng, Jun-Hui and Hofstetter, Walter}}(2019)}]{PhysRevLett.122.010406}%
  \BibitemOpen
  \bibfield  {author} {\bibinfo {author} {\bibnamefont {{Irsigler, Bernhard and Zheng, Jun-Hui and Hofstetter, Walter}}},\ }\bibfield  {title} {\bibinfo {title} {{Interacting Hofstadter Interface}},\ }\href {https://doi.org/10.1103/PhysRevLett.122.010406} {\bibfield  {journal} {\bibinfo  {journal} {{Phys. Rev. Lett.}}\ }\textbf {\bibinfo {volume} {122}},\ \bibinfo {pages} {010406} (\bibinfo {year} {2019})}\BibitemShut {NoStop}%
\bibitem [{\citenamefont {{Goldman, N. and Jotzu, G. and Messer, M. and G\"org, F. and Desbuquois, R. and Esslinger, T.}}(2016)}]{PhysRevA.94.043611}%
  \BibitemOpen
  \bibfield  {author} {\bibinfo {author} {\bibnamefont {{Goldman, N. and Jotzu, G. and Messer, M. and G\"org, F. and Desbuquois, R. and Esslinger, T.}}},\ }\bibfield  {title} {\bibinfo {title} {{Creating topological interfaces and detecting chiral edge modes in a two-dimensional optical lattice}},\ }\href {https://doi.org/10.1103/PhysRevA.94.043611} {\bibfield  {journal} {\bibinfo  {journal} {{Phys. Rev. A}}\ }\textbf {\bibinfo {volume} {94}},\ \bibinfo {pages} {043611} (\bibinfo {year} {2016})}\BibitemShut {NoStop}%
\bibitem [{\citenamefont {{Jaynes, E.T. and Cummings, F.W.}}(1963)}]{1443594}%
  \BibitemOpen
  \bibfield  {author} {\bibinfo {author} {\bibnamefont {{Jaynes, E.T. and Cummings, F.W.}}},\ }\bibfield  {title} {\bibinfo {title} {{Comparison of quantum and semiclassical radiation theories with application to the beam maser}},\ }\href {https://doi.org/10.1109/PROC.1963.1664} {\bibfield  {journal} {\bibinfo  {journal} {{Proceedings of the IEEE}}\ }\textbf {\bibinfo {volume} {51}},\ \bibinfo {pages} {89} (\bibinfo {year} {1963})}\BibitemShut {NoStop}%
\bibitem [{\citenamefont {{Karnaukhov, Igor N.}}(2019)}]{Karnaukhov_2019}%
  \BibitemOpen
  \bibfield  {author} {\bibinfo {author} {\bibnamefont {{Karnaukhov, Igor N.}}},\ }\bibfield  {title} {\bibinfo {title} {{Topological states in the Hofstadter model on a honeycomb lattice}},\ }\href {https://doi.org/10.1016/j.physleta.2019.04.010} {\bibfield  {journal} {\bibinfo  {journal} {{Physics Letters A}}\ }\textbf {\bibinfo {volume} {383}},\ \bibinfo {pages} {2114–2119} (\bibinfo {year} {2019})}\BibitemShut {NoStop}%
\bibitem [{\citenamefont {{Wiegmann, P. B. and Zabrodin, A. V.}}(1994)}]{PhysRevLett.72.1890}%
  \BibitemOpen
  \bibfield  {author} {\bibinfo {author} {\bibnamefont {{Wiegmann, P. B. and Zabrodin, A. V.}}},\ }\bibfield  {title} {\bibinfo {title} {{Bethe-ansatz for the Bloch electron in magnetic field}},\ }\href {https://doi.org/10.1103/PhysRevLett.72.1890} {\bibfield  {journal} {\bibinfo  {journal} {{Phys. Rev. Lett.}}\ }\textbf {\bibinfo {volume} {72}},\ \bibinfo {pages} {1890} (\bibinfo {year} {1994})}\BibitemShut {NoStop}%
\bibitem [{\citenamefont {{Duncan, Callum W. and \"Ohberg, Patrik and Valiente, Manuel}}(2018)}]{PhysRevB.97.195439}%
  \BibitemOpen
  \bibfield  {author} {\bibinfo {author} {\bibnamefont {{Duncan, Callum W. and \"Ohberg, Patrik and Valiente, Manuel}}},\ }\bibfield  {title} {\bibinfo {title} {{Exact edge, bulk, and bound states of finite topological systems}},\ }\href {https://doi.org/10.1103/PhysRevB.97.195439} {\bibfield  {journal} {\bibinfo  {journal} {Phys. Rev. B}\ }\textbf {\bibinfo {volume} {97}},\ \bibinfo {pages} {195439} (\bibinfo {year} {2018})}\BibitemShut {NoStop}%
\bibitem [{\citenamefont {{Gandhi, Shaina and Bandyopadhyay, Jayendra N.}}(2023)}]{PhysRevB.108.014204}%
  \BibitemOpen
  \bibfield  {author} {\bibinfo {author} {\bibnamefont {{Gandhi, Shaina and Bandyopadhyay, Jayendra N.}}},\ }\bibfield  {title} {\bibinfo {title} {{Topological triple phase transition in non-Hermitian quasicrystals with complex asymmetric hopping}},\ }\href {https://doi.org/10.1103/PhysRevB.108.014204} {\bibfield  {journal} {\bibinfo  {journal} {{Phys. Rev. B}}\ }\textbf {\bibinfo {volume} {108}},\ \bibinfo {pages} {014204} (\bibinfo {year} {2023})}\BibitemShut {NoStop}%
\bibitem [{\citenamefont {{Mao, Shijun and Kuramoto, Yoshio and Imura, Ken-Ichiro and Yamakage, Ai}}(2010)}]{doi:10.1143/JPSJ.79.124709}%
  \BibitemOpen
  \bibfield  {author} {\bibinfo {author} {\bibnamefont {{Mao, Shijun and Kuramoto, Yoshio and Imura, Ken-Ichiro and Yamakage, Ai}}},\ }\bibfield  {title} {\bibinfo {title} {{Analytic Theory of Edge Modes in Topological Insulators}},\ }\href {https://doi.org/10.1143/JPSJ.79.124709} {\bibfield  {journal} {\bibinfo  {journal} {{Journal of the Physical Society of Japan}}\ }\textbf {\bibinfo {volume} {79}},\ \bibinfo {pages} {124709} (\bibinfo {year} {2010})}\BibitemShut {NoStop}%
\bibitem [{\citenamefont {{van Dalum, G. A. R. and Ortix, C. and Fritz, L.}}(2021)}]{van_Dalum_2021}%
  \BibitemOpen
  \bibfield  {author} {\bibinfo {author} {\bibnamefont {{van Dalum, G. A. R. and Ortix, C. and Fritz, L.}}},\ }\bibfield  {title} {\bibinfo {title} {{Magnetic impurities along the edge of a quantum spin Hall insulator: Realizing a one-dimensional AIII insulator}},\ }\href {https://doi.org/10.1103/PhysRevB.103.075115} {\bibfield  {journal} {\bibinfo  {journal} {{Phys. Rev. B}}\ }\textbf {\bibinfo {volume} {103}},\ \bibinfo {pages} {075115} (\bibinfo {year} {2021})}\BibitemShut {NoStop}%
\bibitem [{\citenamefont {{McGinley, Max and Cooper, Nigel R.}}(2021)}]{McGinley_2021}%
  \BibitemOpen
  \bibfield  {author} {\bibinfo {author} {\bibnamefont {{McGinley, Max and Cooper, Nigel R.}}},\ }\bibfield  {title} {\bibinfo {title} {{Elastic backscattering of quantum spin Hall edge modes from Coulomb interactions with nonmagnetic impurities}},\ }\href {https://doi.org/10.1103/PhysRevB.103.235164} {\bibfield  {journal} {\bibinfo  {journal} {{Phys. Rev. B}}\ }\textbf {\bibinfo {volume} {103}},\ \bibinfo {pages} {235164} (\bibinfo {year} {2021})}\BibitemShut {NoStop}%
\bibitem [{\citenamefont {{Nathan, Frederik and Abanin, Dmitry and Berg, Erez and Lindner, Netanel H. and Rudner, Mark S.}}(2019)}]{PhysRevB.99.195133}%
  \BibitemOpen
  \bibfield  {author} {\bibinfo {author} {\bibnamefont {{Nathan, Frederik and Abanin, Dmitry and Berg, Erez and Lindner, Netanel H. and Rudner, Mark S.}}},\ }\bibfield  {title} {\bibinfo {title} {{Anomalous Floquet insulators}},\ }\href {https://doi.org/10.1103/PhysRevB.99.195133} {\bibfield  {journal} {\bibinfo  {journal} {{Phys. Rev. B}}\ }\textbf {\bibinfo {volume} {99}},\ \bibinfo {pages} {195133} (\bibinfo {year} {2019})}\BibitemShut {NoStop}%
\bibitem [{\citenamefont {{Titum, Paraj and Berg, Erez and Rudner, Mark S. and Refael, Gil and Lindner, Netanel H.}}(2016)}]{PhysRevX.6.021013}%
  \BibitemOpen
  \bibfield  {author} {\bibinfo {author} {\bibnamefont {{Titum, Paraj and Berg, Erez and Rudner, Mark S. and Refael, Gil and Lindner, Netanel H.}}},\ }\bibfield  {title} {\bibinfo {title} {{Anomalous Floquet-Anderson Insulator as a Nonadiabatic Quantized Charge Pump}},\ }\href {https://doi.org/10.1103/PhysRevX.6.021013} {\bibfield  {journal} {\bibinfo  {journal} {{Phys. Rev. X}}\ }\textbf {\bibinfo {volume} {6}},\ \bibinfo {pages} {021013} (\bibinfo {year} {2016})}\BibitemShut {NoStop}%
\bibitem [{\citenamefont {{Rout, Prasanna and Papadopoulos, Nikos and Pe{\~{n}}aranda, Fernando and Watanabe, Kenji and Taniguchi, Takashi and Prada, Elsa and San-Jose, Pablo and Goswami, Srijit}}(2024)}]{Rout2024}%
  \BibitemOpen
  \bibfield  {author} {\bibinfo {author} {\bibnamefont {{Rout, Prasanna and Papadopoulos, Nikos and Pe{\~{n}}aranda, Fernando and Watanabe, Kenji and Taniguchi, Takashi and Prada, Elsa and San-Jose, Pablo and Goswami, Srijit}}},\ }\bibfield  {title} {\bibinfo {title} {{Supercurrent mediated by helical edge modes in bilayer graphene}},\ }\href {https://doi.org/10.1038/s41467-024-44952-6} {\bibfield  {journal} {\bibinfo  {journal} {{Nature Communications}}\ }\textbf {\bibinfo {volume} {15}},\ \bibinfo {pages} {856} (\bibinfo {year} {2024})}\BibitemShut {NoStop}%
\bibitem [{\citenamefont {{Del Maestro, Adrian and Hyart, Timo and Rosenow, Bernd}}(2013)}]{PhysRevB.87.165440}%
  \BibitemOpen
  \bibfield  {author} {\bibinfo {author} {\bibnamefont {{Del Maestro, Adrian and Hyart, Timo and Rosenow, Bernd}}},\ }\bibfield  {title} {\bibinfo {title} {{Backscattering between helical edge states via dynamic nuclear polarization}},\ }\href {https://doi.org/10.1103/PhysRevB.87.165440} {\bibfield  {journal} {\bibinfo  {journal} {{Phys. Rev. B}}\ }\textbf {\bibinfo {volume} {87}},\ \bibinfo {pages} {165440} (\bibinfo {year} {2013})}\BibitemShut {NoStop}%
\bibitem [{\citenamefont {{Murakami, Shuichi and Nagaosa, Naoto and Zhang, Shou-Cheng}}(2003)}]{doi:10.1126/science.1087128}%
  \BibitemOpen
  \bibfield  {author} {\bibinfo {author} {\bibnamefont {{Murakami, Shuichi and Nagaosa, Naoto and Zhang, Shou-Cheng}}},\ }\bibfield  {title} {\bibinfo {title} {{Dissipationless Quantum Spin Current at Room Temperature}},\ }\href {https://doi.org/10.1126/science.1087128} {\bibfield  {journal} {\bibinfo  {journal} {{Science}}\ }\textbf {\bibinfo {volume} {301}},\ \bibinfo {pages} {1348} (\bibinfo {year} {2003})}\BibitemShut {NoStop}%
\bibitem [{\citenamefont {{Roth, Andreas and Brüne, Christoph and Buhmann, Hartmut and Molenkamp, Laurens W. and Maciejko, Joseph and Qi, Xiao-Liang and Zhang, Shou-Cheng}}(2009)}]{doi:10.1126/science.1174736}%
  \BibitemOpen
  \bibfield  {author} {\bibinfo {author} {\bibnamefont {{Roth, Andreas and Brüne, Christoph and Buhmann, Hartmut and Molenkamp, Laurens W. and Maciejko, Joseph and Qi, Xiao-Liang and Zhang, Shou-Cheng}}},\ }\bibfield  {title} {\bibinfo {title} {{Nonlocal Transport in the Quantum Spin Hall State}},\ }\href {https://doi.org/10.1126/science.1174736} {\bibfield  {journal} {\bibinfo  {journal} {{Science}}\ }\textbf {\bibinfo {volume} {325}},\ \bibinfo {pages} {294} (\bibinfo {year} {2009})}\BibitemShut {NoStop}%
\bibitem [{\citenamefont {{Br{\"u}ne, Christoph and Roth, Andreas and Buhmann, Hartmut and Hankiewicz, Ewelina M. and Molenkamp, Laurens W. and Maciejko, Joseph and Qi, Xiao-Liang and Zhang, Shou-Cheng}}(2012)}]{Brüne2012}%
  \BibitemOpen
  \bibfield  {author} {\bibinfo {author} {\bibnamefont {{Br{\"u}ne, Christoph and Roth, Andreas and Buhmann, Hartmut and Hankiewicz, Ewelina M. and Molenkamp, Laurens W. and Maciejko, Joseph and Qi, Xiao-Liang and Zhang, Shou-Cheng}}},\ }\bibfield  {title} {\bibinfo {title} {{Spin polarization of the quantum spin Hall edge states}},\ }\href {https://doi.org/10.1038/nphys2322} {\bibfield  {journal} {\bibinfo  {journal} {{Nature Physics}}\ }\textbf {\bibinfo {volume} {8}},\ \bibinfo {pages} {485} (\bibinfo {year} {2012})}\BibitemShut {NoStop}%
\bibitem [{\citenamefont {{Anderson, P. W.}}(1958)}]{PhysRev.109.1492}%
  \BibitemOpen
  \bibfield  {author} {\bibinfo {author} {\bibnamefont {{Anderson, P. W.}}},\ }\bibfield  {title} {\bibinfo {title} {{Absence of Diffusion in Certain Random Lattices}},\ }\href {https://doi.org/10.1103/PhysRev.109.1492} {\bibfield  {journal} {\bibinfo  {journal} {{Phys. Rev.}}\ }\textbf {\bibinfo {volume} {109}},\ \bibinfo {pages} {1492} (\bibinfo {year} {1958})}\BibitemShut {NoStop}%
\bibitem [{\citenamefont {{Kramer, B. and MacKinnon, A.}}(1993)}]{Kramer1993}%
  \BibitemOpen
  \bibfield  {author} {\bibinfo {author} {\bibnamefont {{Kramer, B. and MacKinnon, A.}}},\ }\bibfield  {title} {\bibinfo {title} {{Localization: theory and experiment}},\ }\href {https://doi.org/10.1088/0034-4885/56/12/001} {\bibfield  {journal} {\bibinfo  {journal} {{Reports on Progress in Physics}}\ }\textbf {\bibinfo {volume} {56}},\ \bibinfo {pages} {1469} (\bibinfo {year} {1993})}\BibitemShut {NoStop}%
\bibitem [{\citenamefont {{Chuang, Steven and Gao, Qun and Kapadia, Rehan and Ford, Alexandra C. and Guo, Jing and Javey, Ali}}(2013)}]{Chuang2013}%
  \BibitemOpen
  \bibfield  {author} {\bibinfo {author} {\bibnamefont {{Chuang, Steven and Gao, Qun and Kapadia, Rehan and Ford, Alexandra C. and Guo, Jing and Javey, Ali}}},\ }\bibfield  {title} {\bibinfo {title} {{Ballistic InAs Nanowire Transistors}},\ }\href {https://doi.org/10.1021/nl3040674} {\bibfield  {journal} {\bibinfo  {journal} {{Nano Letters}}\ }\textbf {\bibinfo {volume} {13}},\ \bibinfo {pages} {555} (\bibinfo {year} {2013})}\BibitemShut {NoStop}%
\bibitem [{\citenamefont {{Kumar, Mukesh and Nowzari, Ali and Persson, Axel R. and Jeppesen, S{\"o}ren and Wacker, Andreas and Bastard, Gerald and Wallenberg, Reine L. and Capasso, Federico and Maisi, Ville F. and Samuelson, Lars}}(2024)}]{Kumar2024}%
  \BibitemOpen
  \bibfield  {author} {\bibinfo {author} {\bibnamefont {{Kumar, Mukesh and Nowzari, Ali and Persson, Axel R. and Jeppesen, S{\"o}ren and Wacker, Andreas and Bastard, Gerald and Wallenberg, Reine L. and Capasso, Federico and Maisi, Ville F. and Samuelson, Lars}}},\ }\bibfield  {title} {\bibinfo {title} {{Hot Carrier Nanowire Transistors at the Ballistic Limit}},\ }\href {https://doi.org/10.1021/acs.nanolett.4c01197} {\bibfield  {journal} {\bibinfo  {journal} {{Nano Letters}}\ }\textbf {\bibinfo {volume} {24}},\ \bibinfo {pages} {7948} (\bibinfo {year} {2024})}\BibitemShut {NoStop}%
\bibitem [{\citenamefont {{Li, Yiyi and Thompson, Jeff D.}}(2024)}]{PRXQuantum.5.020363}%
  \BibitemOpen
  \bibfield  {author} {\bibinfo {author} {\bibnamefont {{Li, Yiyi and Thompson, Jeff D.}}},\ }\bibfield  {title} {\bibinfo {title} {{High-Rate and High-Fidelity Modular Interconnects between Neutral Atom Quantum Processors}},\ }\href {https://doi.org/10.1103/PRXQuantum.5.020363} {\bibfield  {journal} {\bibinfo  {journal} {{PRX Quantum}}\ }\textbf {\bibinfo {volume} {5}},\ \bibinfo {pages} {020363} (\bibinfo {year} {2024})}\BibitemShut {NoStop}%
\bibitem [{\citenamefont {{Reisenbauer, Manuel and Rudolph, Henning and Egyed, Livia and Hornberger, Klaus and Zasedatelev, Anton V. and Abuzarli, Murad and Stickler, Benjamin A. and Deli{\'{c}}, Uro{\v{s}}}}(2024)}]{Reisenbauer2024}%
  \BibitemOpen
  \bibfield  {author} {\bibinfo {author} {\bibnamefont {{Reisenbauer, Manuel and Rudolph, Henning and Egyed, Livia and Hornberger, Klaus and Zasedatelev, Anton V. and Abuzarli, Murad and Stickler, Benjamin A. and Deli{\'{c}}, Uro{\v{s}}}}},\ }\bibfield  {title} {\bibinfo {title} {{Non-Hermitian dynamics and non-reciprocity of optically coupled nanoparticles}},\ }\href {https://doi.org/10.1038/s41567-024-02589-8} {\bibfield  {journal} {\bibinfo  {journal} {{Nature Physics}}\ }\textbf {\bibinfo {volume} {20}},\ \bibinfo {pages} {1629} (\bibinfo {year} {2024})}\BibitemShut {NoStop}%
\bibitem [{\citenamefont {Bethe}(1997)}]{doi:10.1142/9789812795755_0004}%
  \BibitemOpen
  \bibfield  {author} {\bibinfo {author} {\bibfnamefont {H.}~\bibnamefont {Bethe}},\ }\bibinfo {title} {On the theory of metals, i. eigenvalues and eigenfunctions of a linear chain of atoms},\ in\ \href {https://doi.org/10.1142/9789812795755_0004} {\emph {\bibinfo {booktitle} {Selected Works of Hans A Bethe}}}\ (\bibinfo  {publisher} {{World Scientific}},\ \bibinfo {year} {1997})\ pp.\ \bibinfo {pages} {155--183}\BibitemShut {NoStop}%
\end{thebibliography}%

\end{document}